%% file: paper.tex
\documentclass[epj,final]{svjour}

\usepackage{epsfig}
\usepackage{amsmath}
\usepackage{amssymb}
\usepackage[dvips]{color}
\usepackage{bm}

\newcommand{\lam}{\mbox{$\rm \Lambda$}}
\newcommand{\alam}{\mbox{$\rm \bar \Lambda$}}
\newcommand{\ko}{\mbox{$K^0_S$}}

\begin{document}
\hugehead

\title{Longitudinal Polarization of $\lam$ ~and $\alam$~ Hyperons in Lepton-Nucleon Deep-Inelastic Scattering}

\author{John~Ellis\inst{1}\and Aram~Kotzinian\inst{2,3,4}\and Dmitry~Naumov\inst{2,5} \and Mikhail~Sapozhnikov\inst{2} }

\mail{naumov@nusun.jinr.ru (D.~Naumov)}
\institute{%
Theory Division, Physics Department, CERN, CH~1211~Geneva~23, Switzerland
\and JINR, Dubna, Russia
\and INFN,  Torino, Italy%
\and Yerevan Physics Inst., Yerevan, Armenia%
 \and INFN,  Florence, Italy}

\abstract{%
We consider models for the spin transfers to $\lam$ and $\alam$
hyperons produced in lepton-nucleon deep-inelastic scattering. We
make predictions for longitudinal $\lam$ and $\alam$ spin
transfers for the COMPASS experiment and for HERA, and for the
spin transfer to $\lam$ hyperons produced at JLAB. We demonstrate
that accurate measurements of the spin transfers to $\lam$ and
$\alam$ hyperons with COMPASS kinematics have the potential to
probe the intrinsic strangeness in the nucleon. We show that a
measurement of $\alam$ polarization could provide a clean probe of
the spin transfer from ${\bar s}$ quarks and provides a new
possibility to measure the antistrange quark distribution
function. COMPASS data in a domain of $x$ that has not been
studied previously will provide valuable extra information to fix
models for the nucleon spin structure. The spin transfer to
$\alam$ hyperons, which could be measured by the COMPASS
experiment, would provide a new tool to distinguish between the
SU(6) and Burkardt-Jaffe (BJ) models for baryon spin structure. In
the case of the HERA electron-proton collider experiments with
longitudinally-polarized electrons, the separation between the
target and current fragmentation mechanisms is more clear. It
provides a complementary probe of the strange quark distribution
and helps distinguish between the SU(6) and BJ models for the
$\lam$ and $\alam$ spin structure. Finally, we show that the spin
transfer to $\lam$ hyperons measured in a JLAB experiment would be
dominated by the spin transfer of the intrinsic polarized
strangeness in the remnant nucleon, providing an independent way
to check our model predictions. 
}

\date{Received: date / Revised version:
date~~~~~~~~~~~~~~~~~~~~~~~~~~~~~~~~~~~~~~~~~~~~~~~~~~~~~~~CERN-PH-TH/2007-008}
\markboth{$\lam$ and $\alam$ polarization in lepton nucleon
DIS}{$\lam$ and $\alam$ polarization in DIS}

\maketitle
\input{intro}
\input{model}
\input{results-compass}
\input{conclusions}
\begin{acknowledgement}
We gratefully acknowledge V.Alexakhin, N.Vlasov and Liang Zuo-tang
for useful discussions.
\end{acknowledgement}
\bibliographystyle{aipprocl}
\bibliography{references}
\input{all_figures}
\end{document}

%% file: intro.tex
\section{\label{sec:intro} Introduction}

The origin of the nucleon spin remains a mystery. In a naive non-relativistic constituent
quark model, one might have expected that quarks would carry essentially all the nucleon spin.
 However, measurements of axial-current matrix elements reduced this expectation to 60\% even
 before the first polarized deep-inelastic scattering experiments were performed.
 These experiments have in fact discovered that the quarks carry only about 30\% of the
 nucleon spin. The reasons for this discrepancy are still not clear, and it is also
 uncertain how the rest of the nucleon spin is divided between gluons and orbital
 angular momentum.

One model for the nucleon spin \cite{Bro.88}, motivated by simple
soliton models, is that it is largely due to orbital angular
momentum, with the quark and gluon contribution each small.
Another proposal \cite{Efr.88,Alt.88,Car.88} was that the naive
quark contribution might have been reduced (in one particular
renormalization scheme) by perturbative effects related to the
axial-current anomaly in models with large gluon polarization,
which would need to be counterbalanced by large and negative
orbital angular momentum. A third proposal
\cite{For.89,For.90,For.91} has been that the net quark
contribution might be suppressed by non-perturbative axial-current
anomaly effects. A common feature of all these models is that the
net strange quarks polarization $\Delta s$ is expected to be
negative, for either non-perturbative or (in the case of the
second model) perturbative reasons.

Recent results on high-$p_T$ hadron production from
HERMES~\cite{HERMES-deltaG-old} , SMC~\cite{SMC-deltaG} and
COMPASS ~\cite{COMPASS-high-pt}, charm production data from
COMPASS ~\cite{COMPASS-open-charm} and particle-production
asymmetries from RHIC~\cite{RHIC-deltaG} all indicate that the net
gluon spin, $\Delta G$, is not large enough to suppress
significantly the net quark contribution to the nucleon spin via
the proposed perturbative mechanism. However, the magnitude and
sign of $\Delta G$ are still relatively uncertain
\cite{COMPASS-qcdfit,Leader:2006}, and gluons may still contribute
a large fraction of the proton spin. Neither are the present data
on $\Delta s$ yet conclusive. The data on inclusive polarized
deep-inelastic scattering are interpreted as implying that the net
strange polarization $\Delta s \sim -0.08 \pm
0.02$~\cite{COMPASS-qcdfit,HERMES-g1d}. However, inclusive
experiments provide only indirect evidence for this quantity, and
the HERMES experiment has found no direct evidence for negative
$\Delta s$ in its analysis of particle-production asymmetries in
polarized deep-inelastic scattering~\cite{HERMES-SIDIS}. On the
other hand, these data may not be in a kinematic regime where one
can neglect subasymptotic corrections to the kinematic
distributions, rendering this interpretation
uncertain~\cite{Aram.05}.

An alternative, direct probe of strange quark and antiquark
polarization is provided by the polarization of $\lam$ baryons and
$\bar{\lam}$ antibaryons produced in deep-inelastic scattering.
Their longitudinal polarization is thought to be capable of
`remembering' the spin of the struck strange quark or antiquark,
or the spin of the wounded nucleon remnant left behind in the
target fragmentation region. This proposal \cite{Ell.96} has been
investigated in experiments on the deep-inelastic scattering of
neutrinos, antineutrinos and charged leptons. To date, the most
accurate measurement has been made by the NOMAD experiment, which
found a large negative polarization of $\lam$ baryons, increasing
in the target fragmentation region
\cite{NOMAD_lambda_polar,NOMAD_alambda_polar,Naumov:2002rj,Naumov:2002gs}.
This result is in qualitative agreement with the
polarized-strangeness model \cite{Ell.95,Ell.00}, according to
which a deep-inelastic neutrino collision would scatter
preferentially off a valence quark with positive polarization
along the exchange boson axis, leaving behind a target remnant
containing a negatively polarized strange quark, whose state would
be transferred statistically to $\lam$ baryons produced in the
target fragmentation region. This polarized-strangeness model has
also been used to predict the net $\lam$ polarization in the
COMPASS experiment on polarized muon-nucleon scattering
\cite{EKN}.

The purpose of this paper is to sharpen these predictions for the
COMPASS experiment, to extend them to the \alam~spin
transfer, to make new predictions for the HERA electron-proton collider with
a polarized electron beam,
and to extend the predictions to $\lam$ production in a fixed-target experiment at JLAB.

We show that the polarization of $\alam$ hyperons provides a new
possibility to probe the antistrange quark distribution in the
nucleon. This is because the spin transfer to the $\alam$
depends completely on the antistrange quark distribution, whereas
contributions from the other process to the $\alam$ spin transfer
are negligible, so that the $\alam$ spin transfer
depends only weakly on the associated theoretical uncertainties.
On the other hand, accurate measurements of $\alam$ polarization could provide
valuable information about $\alam$ spin structure. We show that
the different schemes of $\alam$ spin structure result in markedly
different predictions for the $\alam$ polarization, especially at
large $x_F$.

At the HERA energy the target remnant and quark fragmentation
mechanisms may be distinguished unambiguously, making possible
more precise tests of the conjectured net negative polarization of
the $s$ quarks and antiquarks in the nucleon, via studies of
$\lam$ and $\alam$ production in both the remnant and current
fragmentation regions.

 At low energies, for conditions of a typical fixed-target JLAB experiment, we
show that $\lam$ production is dominated by diquark
fragmentation, and extremely sensitive to the correlation between the
polarization of the struck quark and that of the strange quark in the
nucleon remnant.

The processes of spin transfer in the quark fragmentation to
$\lam$ and $\alam$ hyperons have been studied extensively in a
number of theoretical models~
\cite{BJ,Anselmino_group1,Kotzinian_group,Liang_group1,Liang_group2,Dong:2005ea,Liu:2003rk,Liang:2002id,Liu:2000fi,Liang:1997rt,Yang:2002gh,Ma:2001na,Ma:2001rm,Ma:2000cg,Ma:2000uv,deFlorian:1997zj,Ashery:1999am}
for $e^+ e^-$ collisions and lepton nucleon deep-inelastic
scattering processes. In this paper we consider also the
contribution from the target nucleon remnant to the final $\lam$
hyperon polarization.

This paper is organized as follows. In Section~\ref{sec:model}, we
describe the model we use for hadron production in semi-inclusive
deep-inelastic scattering, and describe various scenarios for the
spin transfers to final-state $\lam$ and $\bar{\lam}$ hyperons.
Section~\ref{sec:results-compass} presents results of the
calculations for the different experimental conditions. Finally,
we summarize our conclusions in Section~\ref{sec:conclusion}.

%% file: model.tex
\section{\label{sec:model} Modelling the Hadronic Final State}

We use common definitions for the kinematic variables: $x,y$ are the
standard deep-inelastic scaling variables and $x_F$ is the Feynman
variable defined as
$$x_F=\frac{2P_L^\star}{W},$$
where $P_L^\star$ is the particle longitudinal momentum in the
hadronic centre-of-mass system, whose invariant mass is denoted by
$W$. This definition is approximate and valid only for high W.
However it is commonly used for the lepton-hadron SIDIS in
theoretical and experimental publications.

The negative range of $x_F$ is often referred to as the target
fragmentation region, and the positive range of $x_F$ as the
current fragmentation region. According to QCD factorization
theorems, the hadron production in the current fragmentation
region can be described as a convolution of quark distribution
function in nucleon and quark fragmentation function. Similarly,
one can suppose that particles in the target fragmentation region
are originating from nucleon remnant fragmentation. We note,
however, that the separation between these fragmentation
mechanisms becomes effective only at very high centre-of-mass
energies in lepton-nucleon scattering, so such terminology should
be interpreted with care.

\subsection{Fragmentation Model}

We model the hadronization of quarks and target nucleon remnants
into hadrons within the Lund string model,
which has been used successfully to describe many unpolarized
phenomena, albeit with many free parameters that need to be tuned.
Here we use the Lund model as implemented in the
JETSET~7.4~\cite{JETSET1,JETSET2} code, with the string
fragmentation parameters tuned by the NOMAD
experiment~\cite{ChukanovThesis}.  These parameters describe well
the yields of the strange hadrons $\lam, \alam, \ko$ and - what is
important for our calculations - the relative contributions of
$\lam$ and $\alam$ hyperons produced from decays of heavier states
($\Sigma^*, \Sigma^0, \Xi$ and their
antiparticles)~\cite{NOMAD_lambda_polar,NOMAD_alambda_polar,NOMAD_resonances}.
Deep-inelastic lepton-nucleon scattering is simulated using the
LEPTO~6.1~\cite{LEPTO} package, with parton distributions provided
by the PDFLIB package~\cite{PDFLIB}.

We would like to mention two important improvements to the
LEPTO~6.1 package that we implement here.
\begin{enumerate}
 \item We incorporate lepton scattering off sea $u$ and $d$ quarks,
which was not modelled in the LEPTO code, as was mentioned
in~\cite{EKN}. It is very important because in the original LEPTO
6.1 the scattering over sea $u,d$ quarks was simply treated in the
same way as valence u,d quarks. When scattering occurs on the sea
quark the target remnant is not a simple diquark. Obviously this
changes a lot the final-state hadronization for charged leptons
interactions because of small $x$ dominance at higher energies.
\item In order to describe scattering on a nucleus,
the LEPTO code treats its quark distributions as mixtures of
proton and neutron quark distribution functions weighted by the
relative numbers of protons and neutrons in the nucleus. This has
the feature that, in the event sample so generated, the
distributions of different quark flavours are independent of the
type of target nucleon on which the scattering event takes place,
and also of the final-state hadron selected in semi-inclusive
reactions. We use here a different procedure which describes these
reactions more correctly. For the production of any given hadron,
we simulate separate samples of events produced on individual
proton and neutron targets and combine these samples statistically
to obtain the semi-inclusive cross section for any given hadron on
a composite nuclear target.
The improvement is important because  the $\Lambda$ polarization
strongly depends on the type of nucleon scattered and the quark
scattered. However LEPTO 6.1 mixes the quark density functions of
protons and neutrons if one requires an ``averaged nucleon`` (like
deuteron). This is good procedure for the DIS but not for SIDIS.
For $\Lambda$ production this will give wrong contributions of
heavier states (like $\Sigma^\star, \Sigma^0$ etc). This procedure
also incorporates the correct final-state particle correlations.
\end{enumerate}

\subsection{\label{sec:ranks} Ranks in String Fragmentation}

As preparation for the treatment of spin transfer to final-state
$\lam$ and $\alam$ hyperons, we introduced in~\cite{EKN} two
hadronization ranks $R_q$ and $R_{qq}$.
The ranks $R_q$ and $R_{qq}$ are integers equal to the numerical
ordering of the hadron from the quark and the opposite end, which
we call the nucleon target end~\footnote{For example, for
scattering on a valence quark, this end represents the remnant
diquark of the nucleon, but it can also be a quark or antiquark,
for details see~\cite{LEPTO}.}. Some examples of the assignments
of these two ranks are shown in Fig.~\ref{fig:ranks-definition}.

We often use in the following the terms of quark (target remnant)
fragmentation for the case $R_q < R_{qq}$ ($R_{qq} < R_q$).
However, one must always bear in mind that one is dealing with
string hadronization, and not with independent fragmentation.

\subsection{Models for Spin Transfer}

The polarization of the interacting quark, denoted by $P_q$,
is given by
\begin{equation}
P_q= P_B D(y),
\end{equation}
where $P_B$ is the charged-lepton longitudinal polarization, and
 D(y) is the depolarisation depolarization factor, in the leading
order taken to be $D(y) = \frac{1-(1-y)^2}{1+(1-y)^2}$.
Corrections to $D(y)$ well  parametrized for DIS are poorly known
for SIDIS production of different hadrons.

We assume that there are two basic mechanisms for a baryon to be
produced in a deep-inelastic process with longitudinal
polarization, via spin transfer from the struck quark or from the
target nucleon remnant.
The $\lam$ and $\alam$ hyperons may be produced either
promptly or via the decays of heavier resonances such as
$\Sigma^\star, \Sigma^0, \Xi$ and their antiparticles, which also
transfer partially their polarization to the $\lam$ or $\alam$. We
take both possibilities into account.

The spin transfer from the nucleon target remnant to the $\lam$
and $\alam$ hyperons
can be due to either
polarization of the remnant diquark system (after the struck quark
is removed from the nucleon) or possible sea polarization of
quarks (and antiquarks) produced via intermediate string breaking,
which then fragment into the baryon considered.

Following~\cite{EKN}, we consider in the following two extreme
cases for spin transfer to hyperons.

\begin{itemize}
\item {\bf Model A}: Restrict spin transfer in (di-)quark fragmentation to hyperons with
($R_{qq}=1, R_{q}\neq 1$) $R_{qq}\neq 1, R_{q}=1$.
\item {\bf Model B}: Allow spin transfer in (di-)quark fragmentation to hyperons with
 ($R_{qq}< R_{q}$) $R_{qq} > R_{q}$.
\end{itemize}
The spin transfer in quark and diquark fragmentation is then calculated
as follows.

\subsubsection{Spin Transfer from the Struck Quark}

The polarization of ${\lam}$ and $\alam$ hyperons produced promptly or via
the decay of a strange baryon $Y$ in quark fragmentation is
assumed to be related to the quark
polarization $P_q$ by:
\begin{equation}
\begin{aligned}
\label{eq:cfr}
P_{\lam}^q(Y) = - C_q^{\lam}(Y) P_q,\\
P_{\alam}^{\bar q}(\bar Y) = - C_{\bar q}^{\alam}(\bar Y) P_{\bar q},\\
\end{aligned}
\end{equation}
where $C_q^{\lam}(Y) = C_{\bar q}^{\alam}(\bar Y)$ are the
corresponding spin-transfer coefficients. For the sake of
simplicity we use the notation $C_q^{\lam}(Y)$ for both $\lam$ and
$\alam$.

We use two different models to calculate $C_q^{\lam}(Y)$. The
first one is based on non-relativistic SU(6) wave functions, where
the ${\lam}$ ($\alam$) spin is carried only by its constituent $s$
($\bar s$) quark. In this case, the promptly-produced ${\lam}$
($\alam$) hyperons could be polarized only in $s$ ($\bar s$) quark
fragmentation. The second approach was suggested by Burkardt and
Jaffe (BJ)~\cite{BJ}, who assumed that the modification of the
SU(6) spin decomposition discovered in the case of the nucleon
exists also for other octet hyperons. The spin contents of all
octet baryons within this approach were obtained
in~\cite{Liang:1997rt}. Table~\ref{tab:cfr_coeff} summarizes the
spin-correlation coefficients in the SU(6) and BJ models for both
prompt ${\lam}$ hyperons and octet and decuplet intermediate
hyperons.

\begin{table}[htb]
  \begin{center}
    \caption{\label{tab:cfr_coeff} \it Spin correlation coefficients in the SU(6) and BJ models.}
    \begin{tabular}{c|c|c|c|c|c|c}
      \hline\hline
      ${\lam}$'s parent & \multicolumn{2}{c|}{$C_u^{\lam}$}
      & \multicolumn{2}{c|}{$C_d^{\lam}$}
      & \multicolumn{2}{c}{$C_s^{\lam}$} \\
      \hline\hline
      & SU(6) & BJ & SU(6) & BJ& SU(6) & BJ\\
      \cline{2-7}
      quark              & 0     & -0.18 & 0 & -0.18 & 1 & 0.63\\
      $\Sigma^0$         & -2/9  & -0.12 & -2/9 & -0.12 & 1/9 & 0.15\\
      $\Xi^0$            & -0.15  &  0.07 &  0   & 0.05 & 0.6 & -0.37\\
      $\Xi^-$            &  0    &  0.05 & -0.15 & 0.07 & 0.6 & -0.37\\
      ${\Sigma^\star}$   & 5/9   & -- & 5/9   & -- & 5/9   & -- \\
      \hline\hline
    \end{tabular}
  \end{center}
\end{table}

In Model A the $\lam$ and $\alam$ are polarized according to (\ref{eq:cfr})
 if $R_q=1$ and $R_{qq}\ne 1$. In Model B the corresponding
 condition is $R_q< R_{qq}$. We assume that no spin transfer
 occurs if $R_q = R_{qq}$.

\subsubsection{Spin Transfer from the Remnant Diquark}




We parametrize a possible sea-quark polarization as a {\em
correlation} between the polarization of the sea quark and that of
the struck quark, described by the spin-correlation coefficients
$C_{sq}$:
\begin{equation}
\label{eq:spol}
P_s=C_{sq} P_q,
\end{equation}
where $P_q$ and $P_s$ are the polarizations of the initial struck
quark and the strange quark. The  values of the $C_{sq}$
parameters (one for scattering on a valence quark, the other for
scattering on a sea quark) were found in a fit to NOMAD
data~\cite{EKN}:
\begin{equation}
\label{eq:csq_fit}
\begin{aligned}
\mbox{\bf Model A:} &                                & \\
                    & \hspace{-0.9cm}C_{sq_{val}} = -0.35 \pm 0.05, & \hspace{-0.3cm}C_{sq_{sea}} =-0.95 \pm 0.05.\\
\mbox{\bf Model B:} &                                & \\
                    & \hspace{-0.9cm}C_{sq_{val}} = -0.25 \pm 0.05, & \hspace{-0.4cm}C_{sq_{sea}} = 0.15 \pm 0.05.\\
\end{aligned}
\end{equation}
We calculate  as follows the polarizations of $\lam$ hyperons
produced in target remnant fragmentation:
\begin{equation}
\label{eq:dir}
\begin{aligned}
&P_{\Lambda}^{l\, u}(prompt;N)=P_{\Lambda}^{l\, d}(prompt;N)=C_{sq} \cdot P_q,\\
&P_{\Lambda}^{l \, u}(\Sigma^0;p) =P_{\Lambda}^{l \, d}(\Sigma^0;n) = \frac{1}{3} \cdot \frac{2+C_{sq}}{3+2C_{sq}} \cdot P_q,\\
&P_{\Lambda}^{l \, u}({\Sigma^\star}^0;p) = P_{\Lambda}^{l \, d}({\Sigma^\star}^0;n) = P_{\Lambda}^{l \, d}({\Sigma^\star}^+;p) =\\
&P_{\Lambda}^{l \, u}({\Sigma^\star}^-;n) =-\frac{5}{3} \cdot \frac{1-C_{sq}}{3-C_{sq}} \cdot P_q.
\end{aligned}
\end{equation}
The
equations (\ref{eq:dir}) are applicable only for $\lam$ hyperons,
because we assume that $\alam$ hyperons cannot acquire polarization
from the nucleon remnant.


%% file: results-compass.tex
\section{\label{sec:results-compass} Predictions for Experiments}

For the sake of comparison with experimental data with various
beam polarizations, in what follows we present our results for the
spin transfer as defined by
\begin{equation}
\label{eq:spin_transfer}
S \equiv \frac{P_{\Lambda (\bar\Lambda)}}{P_B D(y)}.
\end{equation}
Most of the calculations presented in this section have been
performed for the COMPASS experimental conditions, namely
deep-inelastic scattering of positive muons with an energy of 160
GeV and beam polarization $P_b=-0.76$ on an isoscalar $^6LiD$
target. We consider scattering on an unpolarized target, as
appropriate for comparison with the first COMPASS experimental
data analysis on spin transfer to $\lam$ and $\alam$
\cite{Sap.06}. We also provide our predictions for two other
experimental cases: HERA $e-p$ collisions ($E_e=27.5$ GeV,
$E_p=820$ GeV) and JLAB fixed-target electron-proton scattering
($E_e=5.7$ GeV). In  both cases, we assume an electron
polarization $P_b=-0.8$. The standard deep-inelastic cut $Q^2>1$
GeV$^2$ was imposed for these experimental conditions.


\subsection{Characteristics of $\lam$ and $\alam$ Production}
In order to understand the spin transfer to hyperons, it is
important first to clarify the relative roles of different
mechanisms for $\lam$ ($\alam$) production in deep-inelastic
scattering.

In Fig.~\ref{fig:xF-distributions} we show the $x_F$ distributions
of $\lam$ and $\alam$ hyperons produced at the COMPASS energy.
The plots are shown for Model B, using the GRV98
parametrization~\cite{GRV98} of the parton distribution functions
and the SU(6) model for the \lam~ spin structure. However, similar
trends are exhibited by Model A and other choices of the parton
distribution functions and \lam~ spin structure.

One may conclude from Fig.~\ref{fig:xF-distributions} (left) that
in all the $x_F$ region the main contribution to $\lam$ production
is that due to diquark fragmentation (dash-dotted line). This
dominance was to be expected in the target fragmentation region
($x_F<0$), but at the COMPASS energy the diquark fragmentation is
important even in the current fragmentation region ($x_F>0$).

As was also demonstrated previously~\cite{EKN}, heavy hyperon
decays are important sources of $\lam$ production. In
Fig.~\ref{fig:xF-distributions} the contribution from heavy
hyperons produced from the target nucleon remnant (dashed line)
and quark fragmentation (dotted line) are shown. The contribution
from the fragmentation of $u$, $d$ light quarks
(dash-triple-dotted line) are considerably smaller than these two
sources, especially at  $x_F<0$. The contribution from the
$s$-quark fragmentation (long dashed line) increases at large and
positive $x_F$. At $x_F=0.5-0.8$ it grows larger than the
contribution from resonance decays, though it is still smaller
than the contribution due to diquark fragmentation.

We now extend the discussion of~\cite{EKN} to $\alam$ production.
Its $x_F$ dependence is shown in
Fig.~\ref{fig:xF-distributions} (right). One can see that diquark
fragmentation (dash-dotted line) is dominant in the region of
small $x_F<0.5$. For $\alam$ production this statement means
simply that the $\alam$ prefers to be formed near the diquark end
of the corresponding string. It differs from the diquark
fragmentation of $\lam$, as in that case the diquark could
fragment directly to the hyperon, and therefore transfer its
polarization to the hyperon. According to our assumptions, there
is no spin transfer in diquark fragmentation to an antihyperon.

We also see that heavy hyperon decays (dashed line) are
insignificant for $\alam$ production, whereas the contribution of
$\bar{s}$-quark fragmentation (long dashed line) is more important
for $\alam$. At large $x_F>0.6$, the $\alam$ are produced mainly
from $\bar{s}$ quarks. This contribution is larger than the
diquark fragmentation, heavy hyperon decays or fragmentation from
light quarks (dash-triple-dotted line). Thus, at the COMPASS
energy, $\alam$ production may serve as a clean source of
information about the antistrange sea of the nucleon, better than
$\lam$ production.

It is interesting that similar dependencies exist at the highest
energy we consider, namely that of HERA. In
Fig.~\ref{fig:xF-distributions_hera} we show the $x_F$
distributions of $\lam$ and $\alam$  produced in DIS with HERA
conditions. One can see that diquark fragmentation (dash-dotted
line) still dominates, even at this HERA energy. The production of
the $\lam$ via heavy hyperon decays (dashed line) is important
only in the region $x_F<0$. The contribution from quark
fragmentation in general, and from the strange quark in
particular, is more clearly separated. The pronounced peak around
$x_F=0$ is due to scattering on sea
light and strange quarks. Even at HERA energy, diquark
fragmentation makes the largest contribution to $\lam$ production,
but conditions for the study of different quark contributions to
the $\lam$ production are better than at lower energy. In the case
of $\alam$ production, we see that the production from ${\bar s}$
fragmentation is dominant at $x_F > 0.5$.

\subsection{Spin Transfer for $\lam$ and $\alam$}

In Fig.~\ref{fig:SpinTansferModelB-Lambdadistributions} we show
the spin transfers (\ref{eq:spin_transfer}) to $\lam$ and $\alam$
hyperons in the SU(6) scheme for model B, as functions of $x_F$ at
the COMPASS energy. The solid line in
Fig.~\ref{fig:SpinTansferModelB-Lambdadistributions} corresponds
to the full calculation,  whereas the amount of spin transfer
without the contribution of the $u,d$ quarks ($\bar{u},\bar{d}$
for $\alam$) is shown as the dotted line, and the spin transfer
without the contribution of the $s$ quarks is shown as the
long-dashed line. The dashed line shows the spin transfer when we
set $C_{sq}=0$ in (\ref{eq:spol}).

The left panel of
Fig.~\ref{fig:SpinTansferModelB-Lambdadistributions} clearly
displays two main mechanisms of spin transfer to $\lam$ hyperons:
target nucleon remnant (at $x_F<0$) and current fragmentation. It
is noteworthy  that switching off the $s$-quark contribution
removes completely  the spin transfer for $x_F > 0.2$. In this
region the polarization  of the $\lam$~ is determined essentially
by the scattering on $s$ quarks. The contribution from the struck
light quarks is small and opposite in sign from the positive spin
transfer from the $s$ quark (mainly due to the $\Sigma^0$
resonance). Switching off the scattering on light quarks almost
removes the spin transfer to $\lam$ hyperons at negative $x_F$
because in this case the production of $\lam$ hyperons from the
target nucleon remnant (and thus spin transfer via $C_{sq}$) is
significantly suppressed.

In the right panel of
Fig.~\ref{fig:SpinTansferModelB-Lambdadistributions} the  spin transfer to $\alam$
hyperons  is shown.
At positive $x_F$ the spin transfer to $\alam$~ is significantly
larger than that to $\lam$. Moreover, {\it the spin transfer to
 the $\alam$~ completely disappears when the $\bar{s}$ contribution
is switched off}. The contribution from the fragmentation of the
light quarks
 is small. Therefore, almost all the spin transfer
 from the lepton to the $\alam$ hyperon is due to fragmentation of the
 antistrange quarks.
 Hence, one may expect that the $\alam$ spin transfer should be
 quite sensitive to the details of the antistrange quark
 distribution in the nucleon.
 In Sect. 3.3-3.6 we will  study just to what extent this feature depends
 on the models considered.

Whilst the $x_F$ distributions for the spin transfers to $\lam$ and
$\alam$ hyperons reflect largely the main spin-transfer
mechanisms, we have also to keep in mind the $x$ dependence of the spin
transfer to $\lam$ hyperons due to the different $C_{sq}$ coefficients
(\ref{eq:csq_fit}). Fig.~\ref{fig:SpinTansferModelB-2D} displays the
spin transfers to $\lam$ and $\alam$ hyperons for the COMPASS
conditions as functions of $x$ and $x_F$. In the left panel of
Fig.~\ref{fig:SpinTansferModelB-2D} there are two visible
regions with significant spin transfer to $\lam$ hyperons.
The first region is where $x_F < 0$, and  the spin transfer from the
target nucleon remnant is dominant, whereas the second region is where $x_F >
0$, and $s$-quark fragmentation dominates at small $x$.

Examining  Fig.~\ref{fig:SpinTansferModelB-2D} for the $\lam$
hyperon case, one can see that the $x$ dependence of the spin
transfer is different in different slices of the $x_F$ variable,
especially comparing the domains $x_F<0$ and $x_F>0$. One
must take this into account when comparing theory
predictions with experimental data folded with real experimental
acceptance.

In the right panel of Fig.~\ref{fig:SpinTansferModelB-2D} we
show the spin transfer to $\alam$ hyperons, which is mainly due to
$\bar s$ fragmentation and has a simpler structure. This
facilitates the investigation of the $\bar s$ quark distribution using the
spin transfer to $\alam$.

In Fig.~\ref{fig:SpinTansferModelB-compare_lambda_antilambda} we
display  the spin transfers to $\lam$ and $\alam$ hyperons as
functions of $x$ and $y$. The calculations are made in the $SU(6)$
model with the GRV98 parton density functions. We see that the
model predicts small (or zero) spin transfer to $\lam$ in the
region $x \sim 10^{-2}$, which increases monotonously to $S \sim
0.1$ at $x \sim 10^{-1}$. The $x$ dependence of the spin transfer
to $\alam$ is different from that of the $\lam$, being quite
constant in $x$ around $S \sim 0.05$. The exact value of $S$
depends on the theoretical option used. For instance, it is
different in Models A and B, providing a good test for
discriminating between them.

 The $y$ dependencies of the spin transfers to $\lam$ and $\alam$ are
also different, as shown in Fig.~\ref{fig:SpinTansferModelB-compare_lambda_antilambda} (right)
for Model B. The $y$ dependence of the spin transfer to $\alam$
is quite constant, as it should be if the quark fragmentation
mechanism of the $\alam$ production dominates. On the other hand, the $y$ dependence
of the spin transfer to $\lam$ decreases with $y$, reflecting the
role of the target remnant. It is
 interesting that dropping the contribution from the target remnant, i.e.,
 assuming  $C_{sq}=0$ (or considering only the region $x_F>0$), leads to the disappearance of the $y$
 dependence for $\lam$. It becomes almost constant with a spin-transfer value that is about a factor of two smaller.

\subsection{Comparison between Models A and B}

In Fig.~\ref{fig:SpinTansferModelsAB} we compare the predictions of
Models A and B for the $x_F$, $x$ and $y$ distributions of the spin
transfer to $\lam$ and $\alam$ hyperons in the COMPASS experimental
conditions.

It is worthwhile recalling that the free parameters of our models
for the spin transfer from the intrinsic strangeness of the
nucleon ($C_{sq}^{val},C_{sq}^{sea}$) were tuned using data from
the NOMAD experiment~\cite{NOMAD_lambda_polar}. The COMPASS
kinematics extends the $x$ region to much lower values not
accessible to NOMAD, where the two Models A and B differ
significantly in their predictions. We recall also that the fit to
the NOMAD data performed in~\cite{EKN} yielded negative
correlation coefficients $C_{sq_{val}}, C_{sq_{sea}}$ for Model A,
but negative $C_{sq_{val}}$ and positive $C_{sq_{sea}}$ for Model
B, as can been seen in  (\ref{eq:csq_fit}). Therefore, the two
models diverge in their predictions at small $x$ where the role of
sea quarks is significant, as can be seen from the $x$ dependence
of the spin transfer to $\lam$ hyperons shown in
Fig.\ref{fig:SpinTansferModelsAB}. This is reflected in the large
difference between models A and B at negative $x_F$. A  larger
contribution to the spin transfer to $\lam$ hyperons from
resonances produced in the quark fragmentation is generally found
in Model B. This is reflected in a smaller spin transfer at
positive $x_F$ in Model B, as compared to Model A.

Remarkably, in the case of $\alam$ polarization the predictions of
Models A and B are nearly the same. This stability is due in part
to a compensation of the primary spin transfer to the $\alam$
hyperons by antiresonance contributions. On the one hand, Model B
allows more $\alam$ to be produced from quark fragmentation (and
thus a higher spin transfer is expected on average), on the other
hand it also increases the fraction of $\alam$ hyperons produced
from antiresonance decays. These are dominated by $\bar\Sigma^0$
decays that transfer the opposite polarization.

Thus, the $\lam$ data could be used to discriminate between Models
A and B. On the other hand, the predictions of the spin transfer
to $\alam$ are robust and do not depend on the assumption about
the spin-transfer assignment in a certain ranking scheme.

\subsection{\label{sec:sensitivity_parton}Sensitivity to the Strange Quark Parton Distributions}

 In Fig.~\ref{fig:cteq} we show the spin transfers to the
$\lam$ and $\alam$ hyperons as calculated using the CTEQ5L set of
parton distribution functions~\cite{CTEQ} for the COMPASS energy,
enabling us to demonstrate the effects of different strange quark
parton distributions. The GRV98 set is based on the assumption
that there is no intrinsic nucleon strangeness and the strange sea
is of pure perturbative origin. On the other hand, the CTEQ
collaboration takes into account the dimuon data of the CCFR and
NuTeV experiments~\cite{CCFR,NuTeV}. As a result, the $s(x)$
distribution of CCFR is larger than the GRV98 one by a factor of
about two in the region $x=0.001-0.01$.
To show how this difference in $x$-dependence reflected in the
$x_F$ dependence we plot in Fig.~\ref{fig:cteq} the spin transfers
to the $\lam$ and $\alam$ hyperons. It turns out that the CTEQ
parton distribution results in larger spin transfer to both $\lam$
and $\alam$.

Therefore, precise measurements of the $\alam$ spin transfer could
give useful information about the antistrange quark distribution,
which could be complementary to the other inclusive deep-inelastic data,
which is sensitive to the unpolarized sum $s(x)+\bar{s}(x)$.

In Fig.~\ref{fig:compare_grv98_cteq5l_su6_hera} we display our
predictions for the spin transfer to the $\lam$ and $\alam$
hyperons for the GRV98 and CTEQ5L sets  of parton distributions as
functions of $x_F$ with HERA kinematics.
 These plots show an interesting
effect appearing at high center-of-mass energies, namely the
spin transfer to the $\alam$ decreases its sensitivity to the parton
 distributions.
 In fact, the spin transfer
to $\lam$ hyperons shows a similar trend as in case of the COMPASS
kinematics with a lower center-of-mass energy (see
Fig.~\ref{fig:cteq}), while $\alam$ hyperons are apparently not as
sensitive as one might naively expect.  This effect can be
understood as follows. At high center-of-mass energies the charged
leptons scatter preferentially on sea quarks and antiquarks, and
quite often the hadronic string is produced between a quark and
antiquark pair from the nucleon sea (in contrast to the case of
lower center-of-mass energy, where the hadronic string is
typically produced between a scattered valence quark and remnant
diquark). This explains the peak around zero in the $x_F$
distribution of the produced hadrons (as seen in
Fig.~\ref{fig:xF-distributions_hera}). The majority of $\alam$
hyperons is produced from the fragmentation of the color string
stretched between a $u-\bar u$ (or $d-\bar d$) pair with an
increasing contribution from $s-\bar s$ at larger $x_F$. However
even those produced from the $s-\bar s$ color string fragmentation
often are far from the struck $\bar s$ end and thus are not
polarized. This explains the loss of sensitivity of spin transfer
to $\alam$ hyperons to the $\bar s(x)$ distribution at higher
center-of-mass energies. On the other hand, the spin transfer to
$\lam$ hyperons is still sensitive to the $s(x)$ distribution
because there is an additional mechanism for producing $\lam$
hyperons from the string stretched from the struck $s$ quark
target nucleon remnant (just as a $s -ud$ string may produce a
$K^+$ or $K^{\star +}$ meson without fragmentation from $u\bar s$
remnant quarks).

\subsection{Comparison between the SU(6) and BJ Models for $\lam$ Spin Structure}

We demonstrate now that accurate measurement of $\alam$ polarization
could provide valuable information about $\alam$ spin structure.

In Fig.~\ref{fig:su6-bj}  the spin transfers to $\lam$ and $\alam$
hyperons in the SU(6) (solid line) and BJ (dashed line) models as
functions of $x_F$ are shown for the case of COMPASS kinematics. A
similar study is displayed in Fig.~\ref{fig:su6-bj-hera} for the
case of HERA kinematics. It is apparent that the $x_F>0$ domain
provides a clean tool for discriminating between the SU(6) and BJ
models for both $\lam$ and $\alam$ and for both COMPASS and HERA.
On the other hand, as we noticed in
Sec.~\ref{sec:sensitivity_parton}, the HERA case for $\alam$ is
not very sensitive to the $\bar s$ parametrization, so a
measurement of the spin transfer to $\alam$ at HERA would provide
a complementary tool for separating the SU(6) and BJ models.

\subsection{$\lam$ Spin Transfer and the Sea-Quark Polarization}

We have demonstrated that at high energies the $\alam$ spin
transfer is sensitive to the fragmentation of the antistrange
quark. As we now show, it is also interesting that at low energies
the $\lam$ spin transfer could provide important information about the
polarization of the sea quarks.

We have calculated the $\lam$ spin transfer for DIS ($Q^2>1~GeV^2$) in  typical JLAB conditions,
assuming $E_e = 5.7$~GeV.

At this energy  the diquark contribution to $\lam$ production
dominates all the other possibilities, for all values of $x_F$.
Therefore, in this case spin-transfer measurements offer an ideal
opportunity to probe the diquark mechanism for spin transfer in
which the polarization of the target remnant is due to the
correlation between the polarizations of the struck quark and the
sea quark defined in (\ref{eq:spol}).

We show in Fig.~\ref{fig:su6_grv98_jlab} the predictions of our
model for the spin transfer to the $\lam$ hyperons in the SU(6)
models for the GRV98 set of parton distributions, as functions of
$x$ and $x_F$. The solid line corresponds to the full calculation,
whereas the amount of spin transfer without the contribution of
the $u,d$ quarks  is shown as the thin dashed line, and the spin
transfer without the contribution of the $s$ quarks is shown as
the dash-dotted line. The bold dashed line shows the spin transfer
when we set $C_{sq}=0$ in (\ref{eq:spol}).
 One can  see that the spin transfer is not
sensitive to the strange or light quark contributions. The main
important contribution is due to strange sea-quark polarization.
Switching off this contribution, i.e., assuming that $C_{sq}=0$,
leads to a significant change in the behaviour of the spin
transfer to $\lam$ .

%% file: conclusions.tex
\section{\label{sec:conclusion} Conclusions}

We introduced in our previous work~\cite{EKN} theoretical and
phenomenological studies of the spin transfers to $\lam$ hyperons
from various mechanisms in lepton-nucleon deep-inelastic
scattering and presented model that described adequately all the
data available at that time on $\lam$ polarization in (anti-)
neutrino-nucleon and charged lepton-nucleon deep-inelastic
scattering. The model had only two free parameters, which were
fitted from NOMAD data~\cite{NOMAD_lambda_polar}. In this paper we
have extended and sharpened our predictions, introducing also the
consideration of the spin transfer to $\alam$ hyperons produced in
various deep-inelastic scattering processes. Our main predictions
have been for the spin transfer to $\lam$ and $\alam$ hyperons in
the COMPASS experiment, which is taking data at the moment, and we
have also provided predictions for the ongoing HERA
electron-proton collider experiments H1 and ZEUS with
longitudinally-polarized electrons, and for the spin transfer to
$\lam$ hyperons in a JLAB fixed-target experiment.

We have demonstrated that the accurate measurement of the spin transfers to $\lam$ and
$\alam$ hyperons with COMPASS kinematics
has the potential to probe the
intrinsic strangeness in the nucleon. We have shown that a
measurement of $\alam$ polarization could provide a clean probe of
the spin transfer from ${\bar s}$ antiquarks. It provides a new
possibility to measure the antistrange quark distribution
function.

Finally, our results for a JLAB fixed target experiment indicate
that spin transfer from the polarized electron beam to $\lam$
hyperons, even produced on unpolarized proton target, would be
dominated by the intrinsic polarized strangeness in the nucleon.
Therefore, a measurement of $\lam$ polarization at JLAB would
provide a new cross-check of polarized intrinsic strangeness in
the nucleon.

These studies have shown that the experimental opportunities
available to COMPASS, HERA and JLAB are largely complementary. In
combination, they could provide valuable information about
polarization of the strange quarks in the nucleon.

%% file: all_figures.tex
\section*{Figures}
\begin{figure}[htb]
\epsfig{file=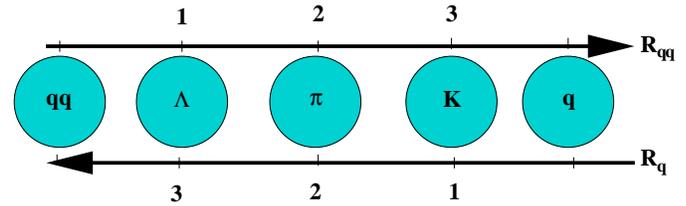,width=\linewidth}
\caption{\label{fig:ranks-definition} The definitions of the two
ranks $R_q$ and $R_{qq}$ that represent the numerical orderings
from the  quark and diquark ends of the Lund fragmentation string,
respectively. In the example shown in this figure the $\lam$
hyperon has $R_{qq}=1$ and $R_q=3$, whereas the $\pi$ has
$R_{qq}=R_q= 2$.}
\end{figure}
\begin{figure*}[htb]
\begin{tabular}{cc}
\epsfig{file=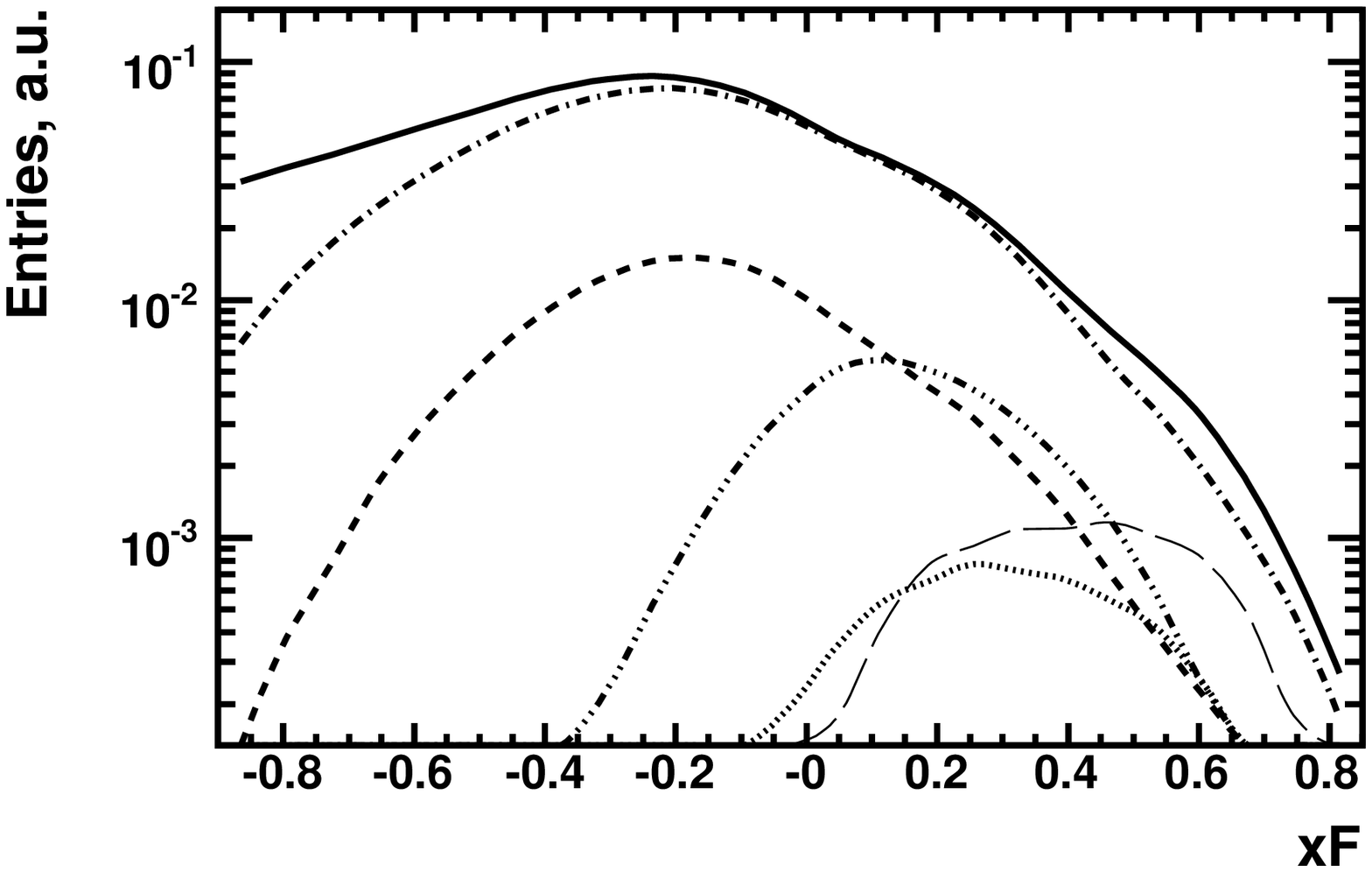,width=0.45\linewidth} &
\epsfig{file=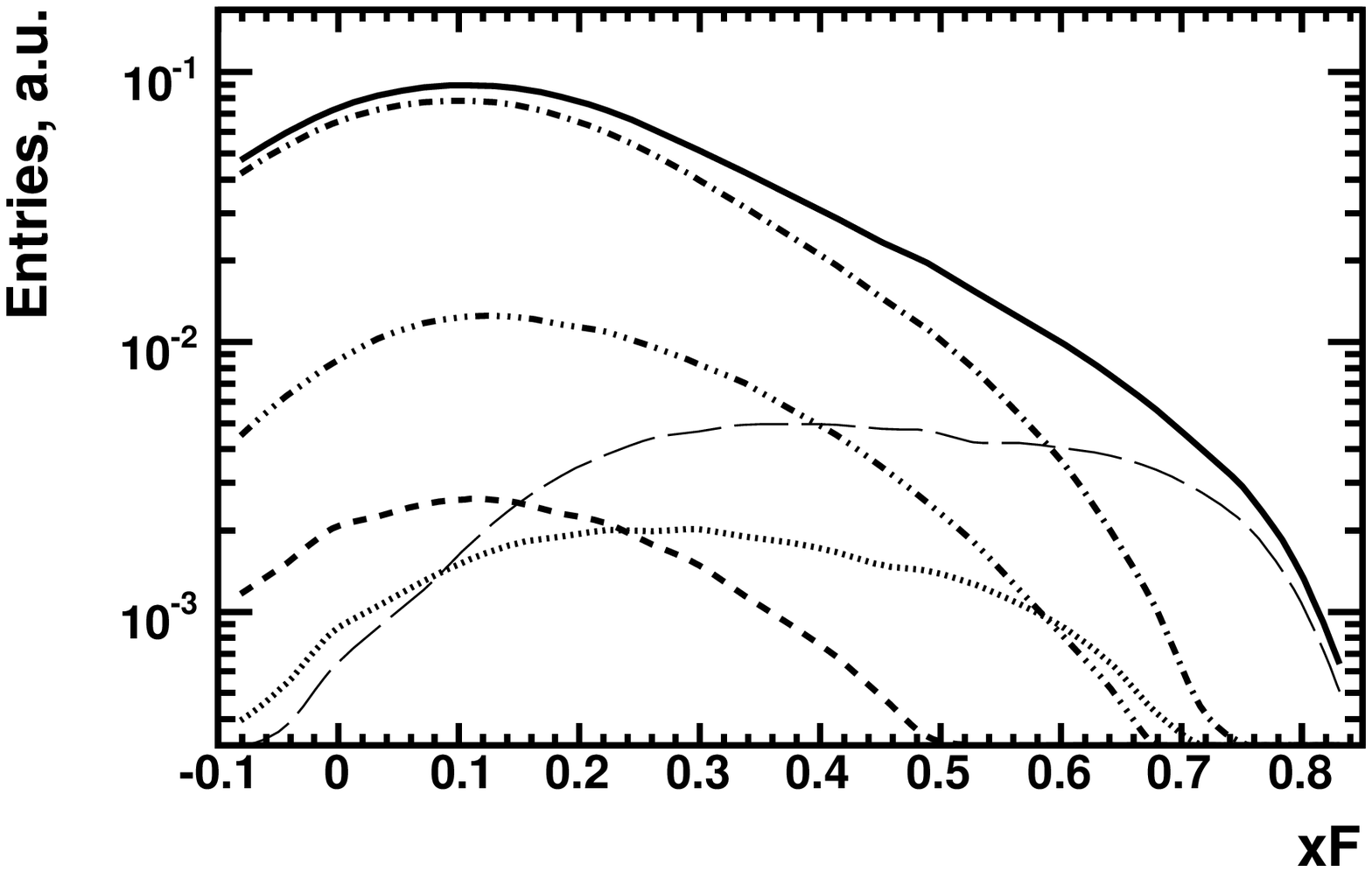,width=0.45\linewidth} \\
\end{tabular}
\caption{\label{fig:xF-distributions} Normalized $x_F$
distributions of $\lam$ (left) and $\alam$ (right) produced via
different channels:
long-dashed line - from $s$ ($\bar{s}$),
dash-triple-dotted line  - from $u$ and $d$ ($\bar{d}$, $\bar{u}$)
light quarks,
dash-dotted line - from the target nucleon end,
dashed line - from decays of heavier resonances produced from the
nucleon end of the string,
 dotted line - from decays of heavier
resonances produced by quark fragmentation.
 These
calculations were performed for the COMPASS energy in Model B
using the GRV98 parametrization of the parton distribution
functions and the SU(6) model of the baryon spin structure.}
\end{figure*}
\begin{figure*}[htb]
\begin{tabular}{cc}
\epsfig{file=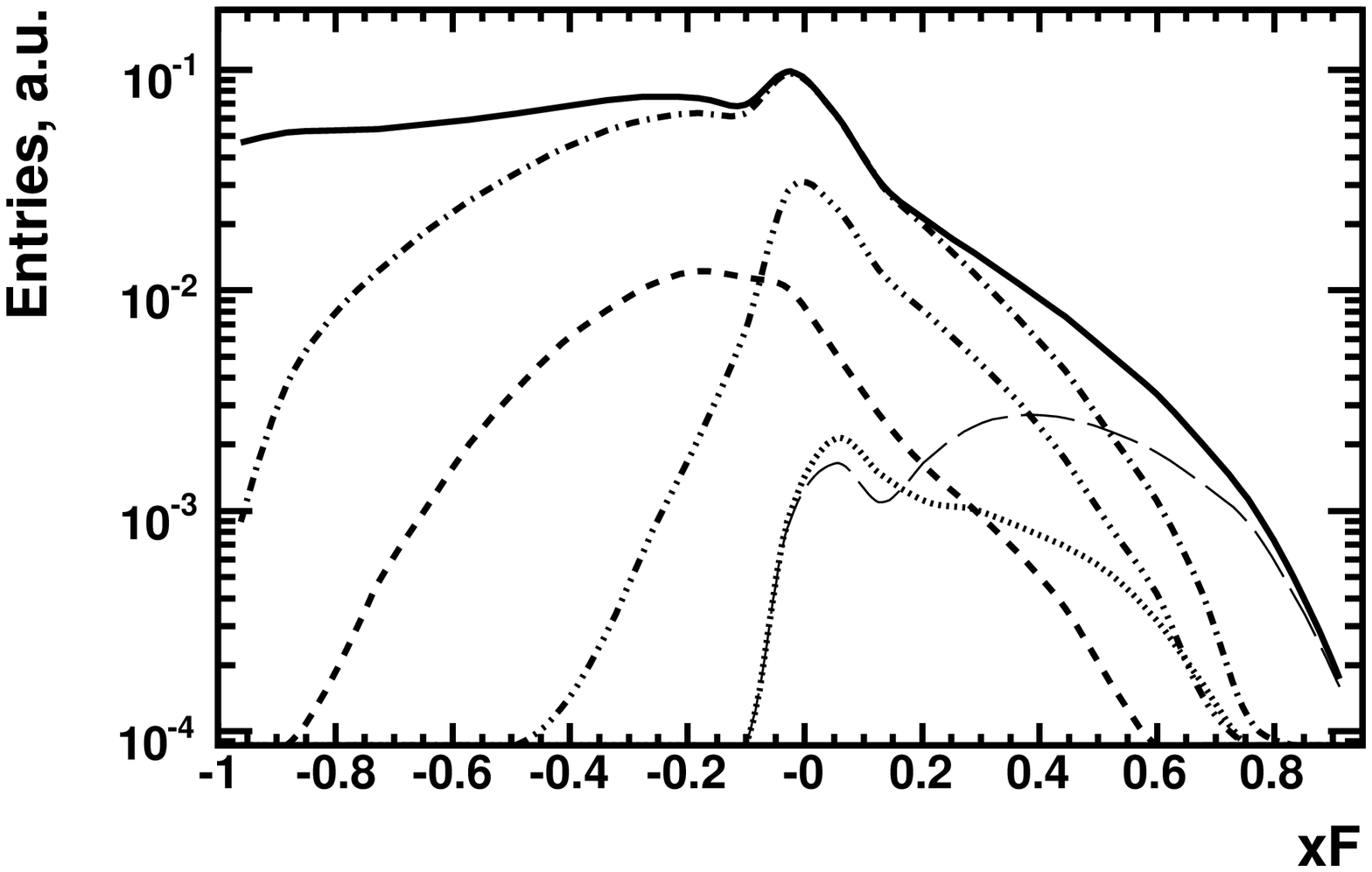,width=0.45\linewidth} &
\epsfig{file=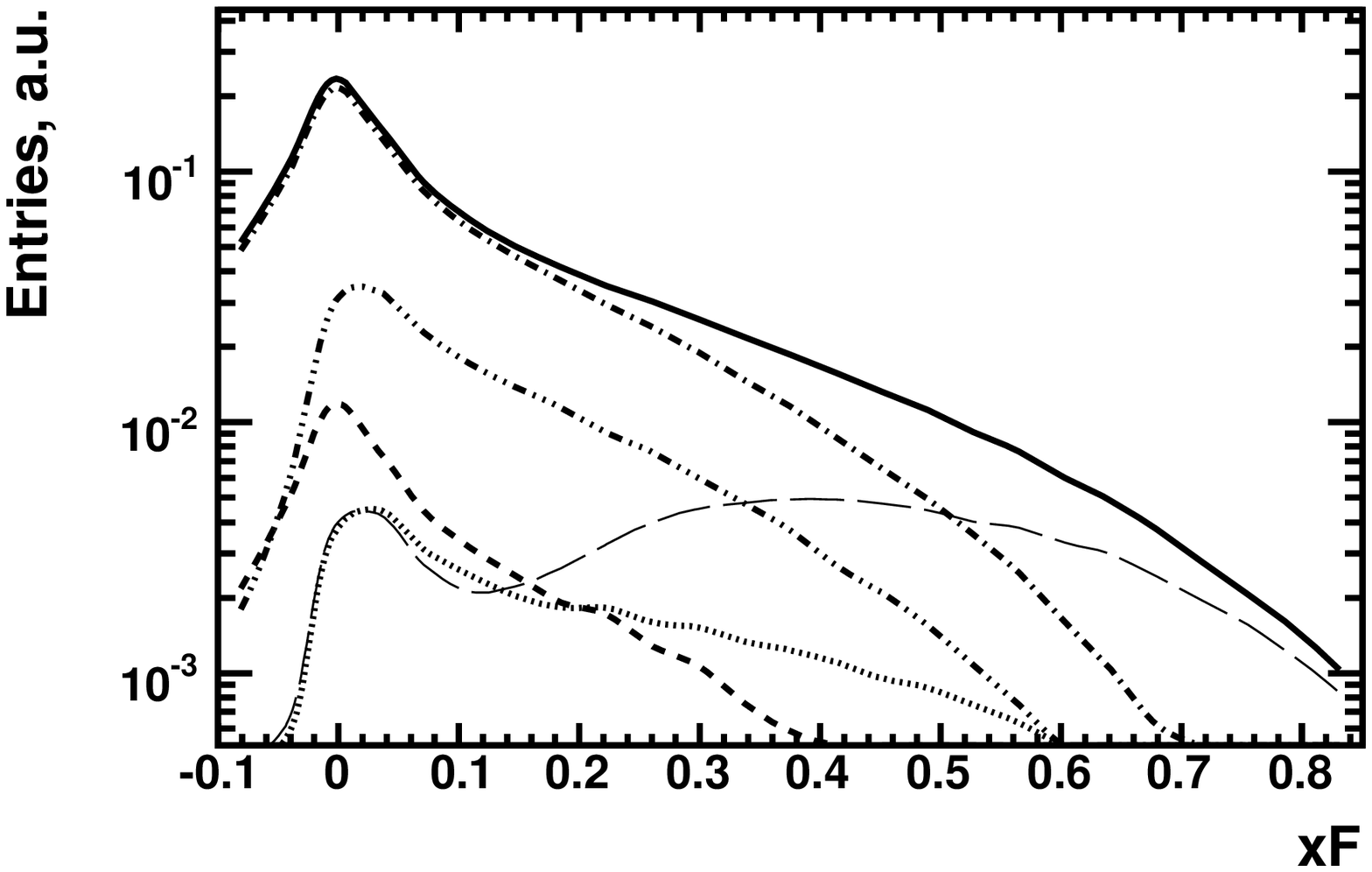,width=0.45\linewidth} \\
\end{tabular}
\caption{\label{fig:xF-distributions_hera} Normalized $x_F$
distributions of the $\lam$ (left) and $\alam$ (right) produced at
HERA energy via different channels: long-dashed line - from $s$
($\bar{s}$), dash-triple-dotted line  - from $u$ and $d$
($\bar{d}$, $\bar{u}$) light quarks, dash-dotted line - from the
target nucleon end, dashed line - from decays of heavier
resonances produced from the nucleon end of the string,
 dotted line - from decays of heavier
resonances produced by quark fragmentation.
 These
calculations were performed in Model B using the GRV98
parametrization of the parton distribution functions and the SU(6)
model of the baryon spin structure.}
\end{figure*}

\begin{figure*}[htb]
\begin{tabular}{cc}
\epsfig{file=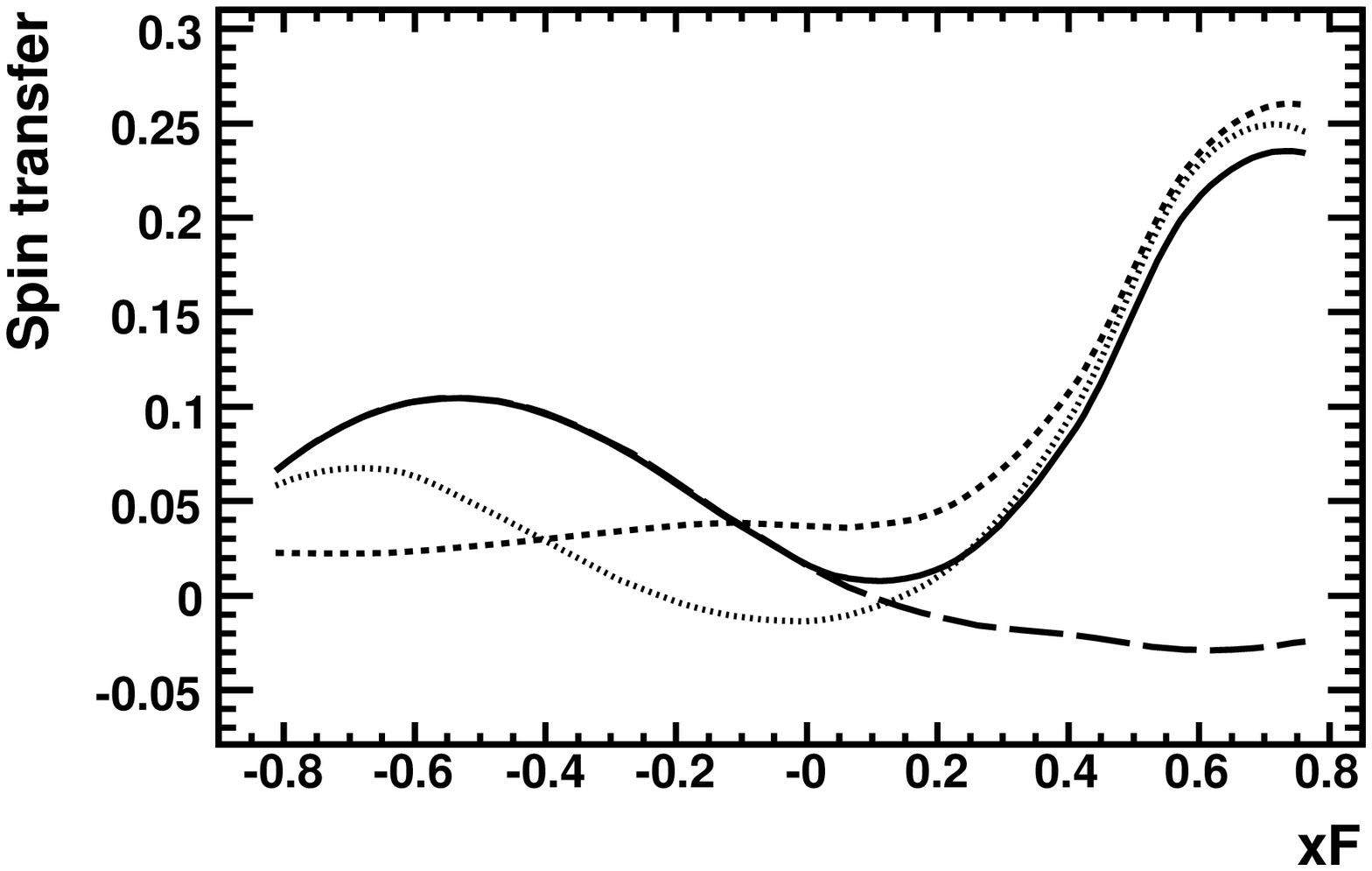,width=0.45\linewidth} &
\epsfig{file=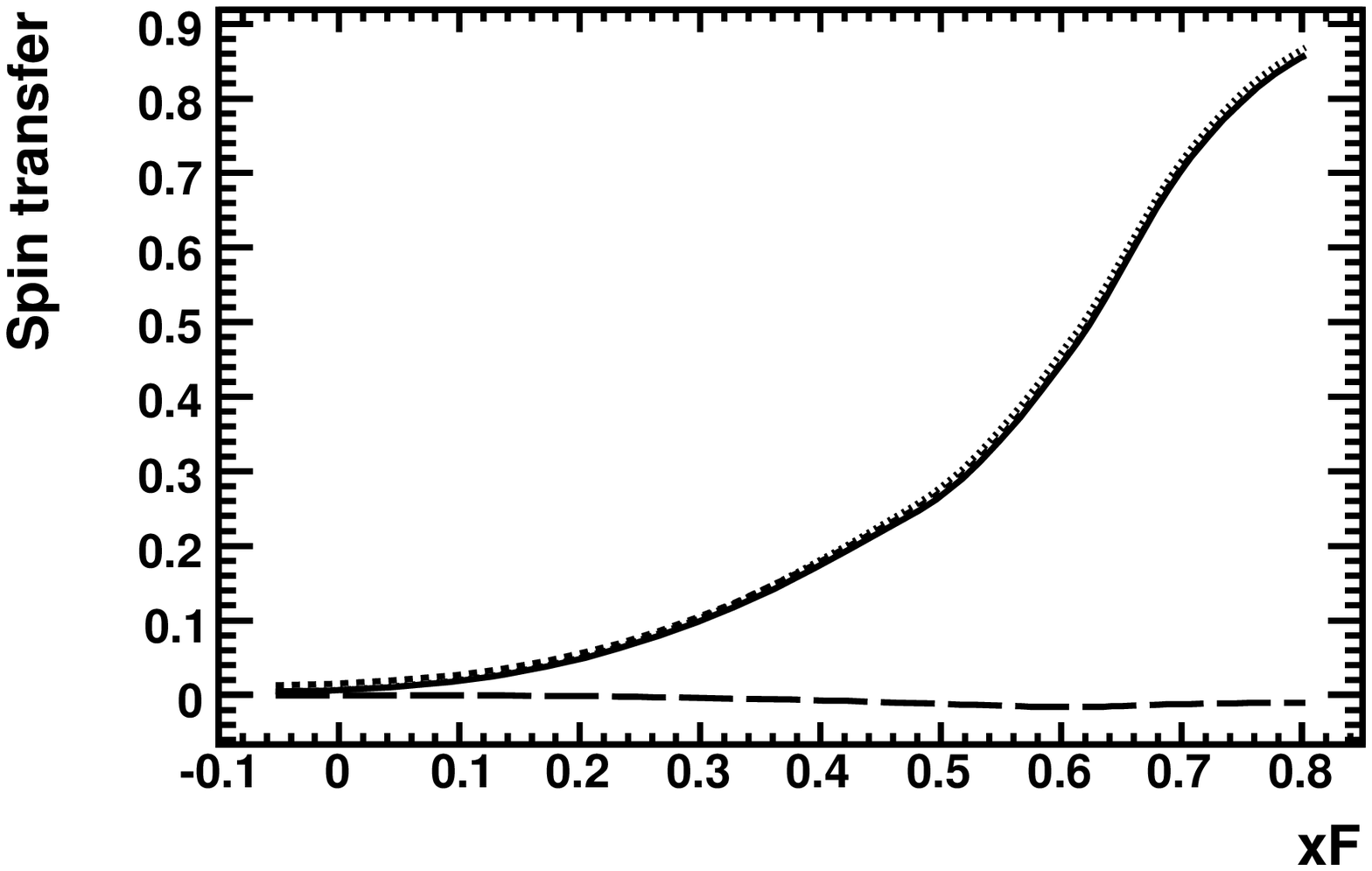,width=0.45\linewidth} \\
\end{tabular}
\caption{\label{fig:SpinTansferModelB-Lambdadistributions} The
spin transfer to $\lam$ hyperons (left) and $\alam$ hyperons
(right) in the SU(6) variant of Model B, as a function of $x_F$,
at COMPASS energy. The solid lines correspond to the full
calculations, the spin transfers without the contributions from
$u$ and $d$ quarks are shown by dotted lines, the spin transfers
without the contributions from the $s$ ($\bar s$) struck quarks
are shown by the long-dashed lines, and the dashed lines
correspond to calculations with $C_{sq}=0$.}
\end{figure*}
\begin{figure*}[htb]
\begin{tabular}{cc}
\epsfig{file=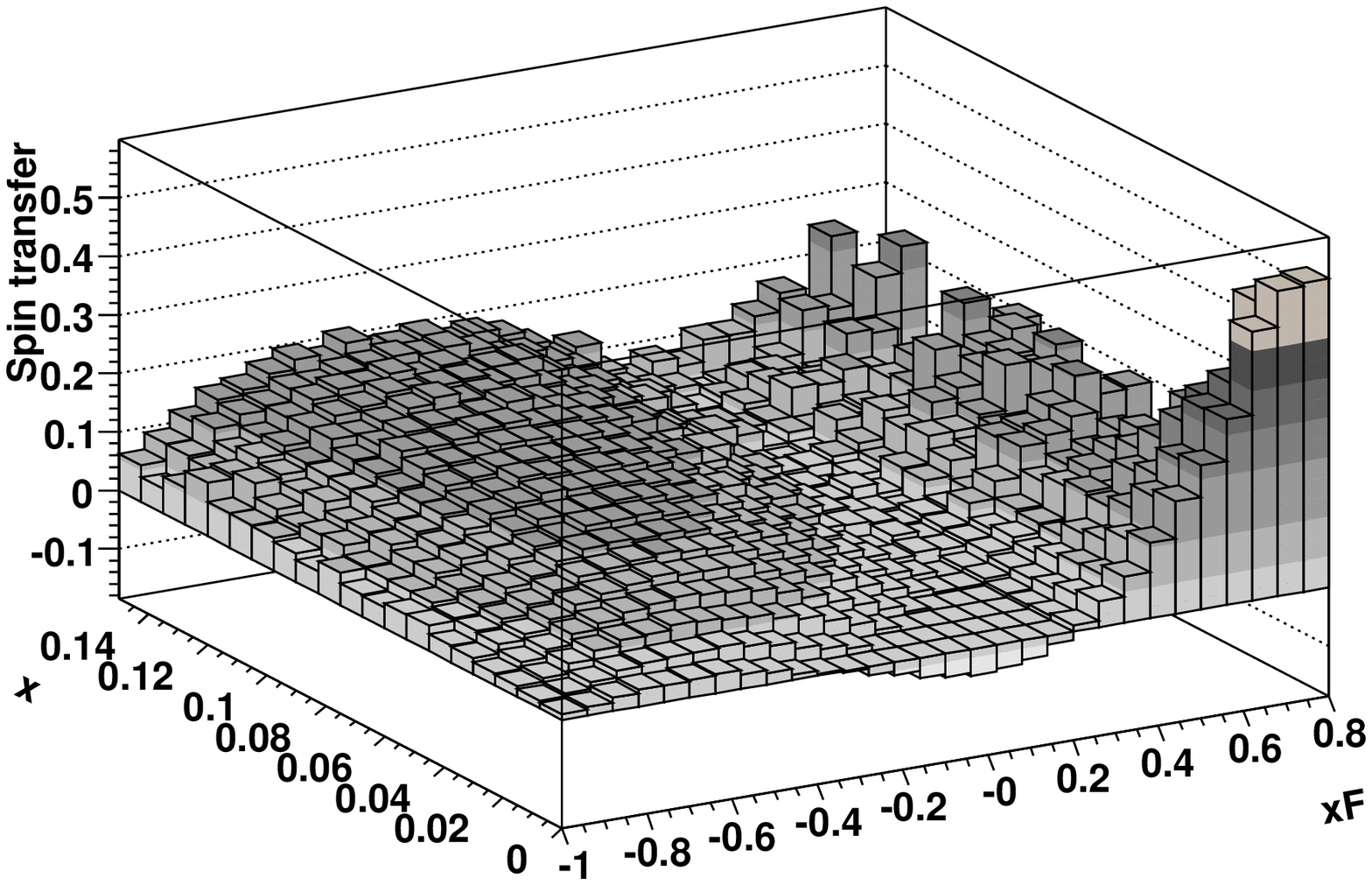,width=0.45\linewidth} &
\epsfig{file=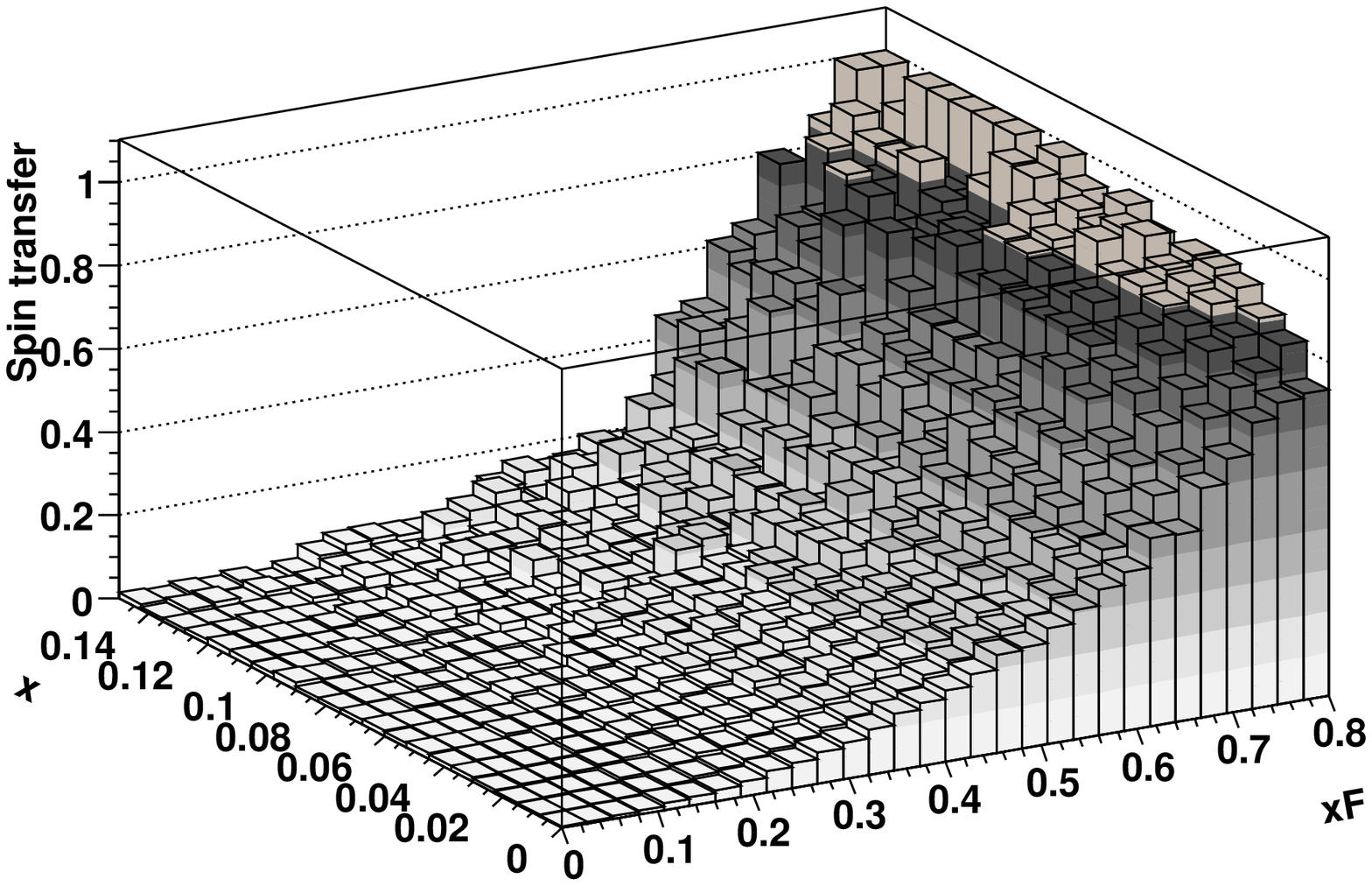,width=0.45\linewidth} \\
\end{tabular}
\caption{\label{fig:SpinTansferModelB-2D} The spin transfers to
$\lam$  (left) and $\alam$ hyperons (right) in the SU(6) variant
of Model B, as functions of $x_F$ and $x$, at COMPASS energy. }
\end{figure*}
\begin{figure*}[htb]
\begin{tabular}{cc}
\epsfig{file=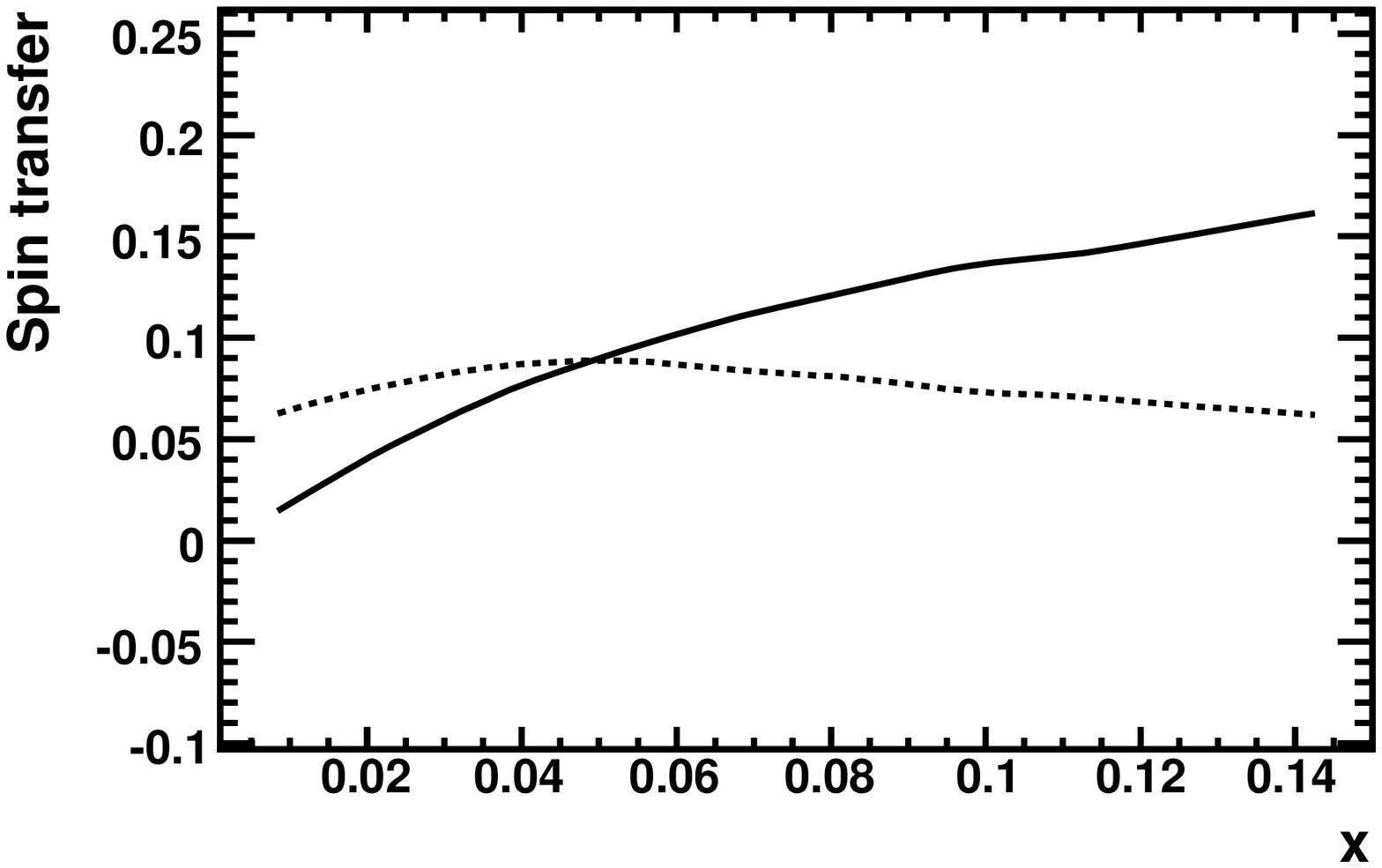,width=0.45\linewidth} &
\epsfig{file=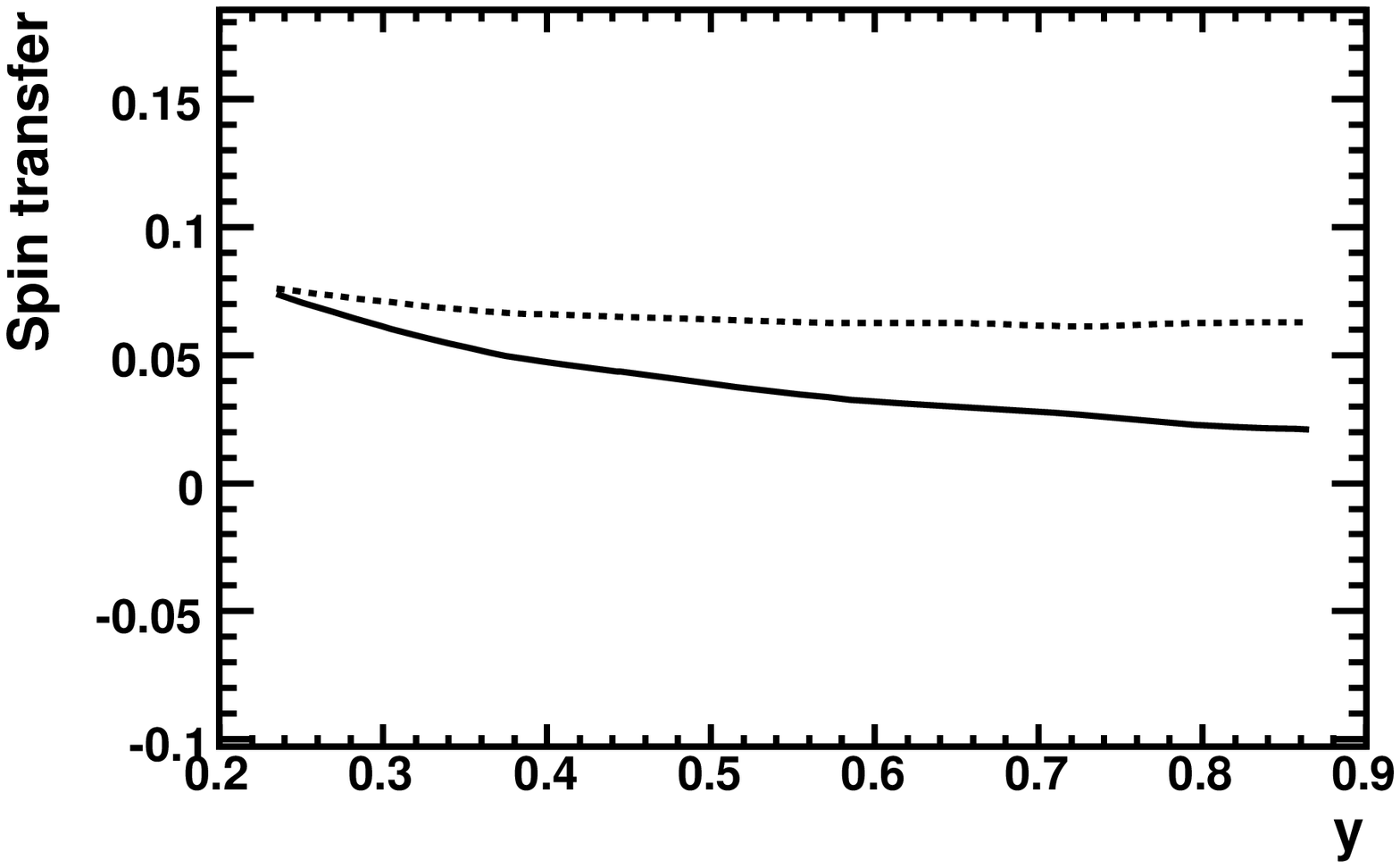,width=0.45\linewidth} \\
\end{tabular}
\caption{\label{fig:SpinTansferModelB-compare_lambda_antilambda}
The spin transfers to $\lam$ (solid line) and $\alam$ (dashed
line) hyperons in the SU(6) variant of Model B, as functions of
$x$ and $y$, at COMPASS energy. }
\end{figure*}

%
%
\begin{figure*}[htb]
\begin{tabular}{cc}
\epsfig{file=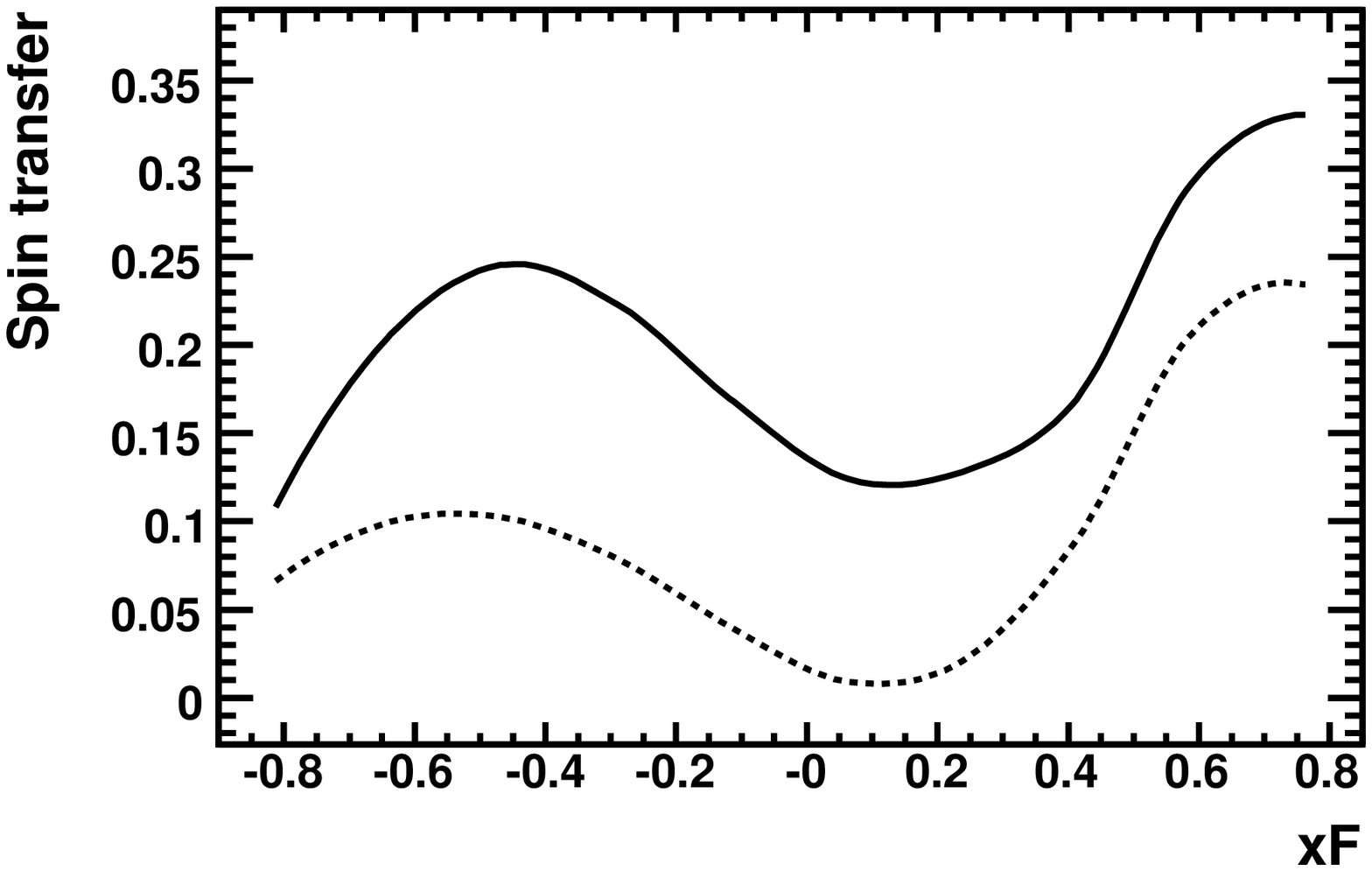,width=0.45\linewidth} &
\epsfig{file=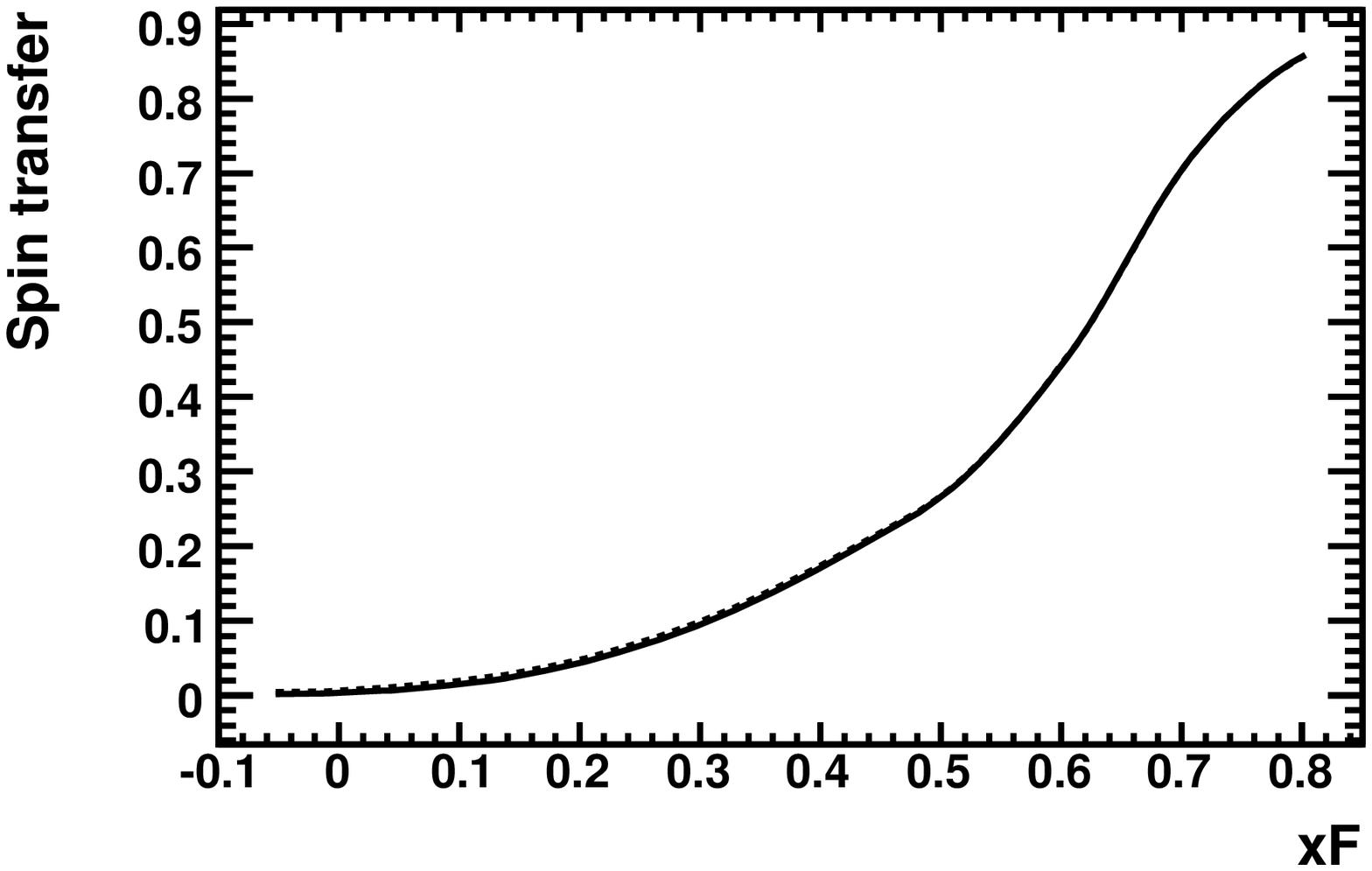,width=0.45\linewidth} \\
\epsfig{file=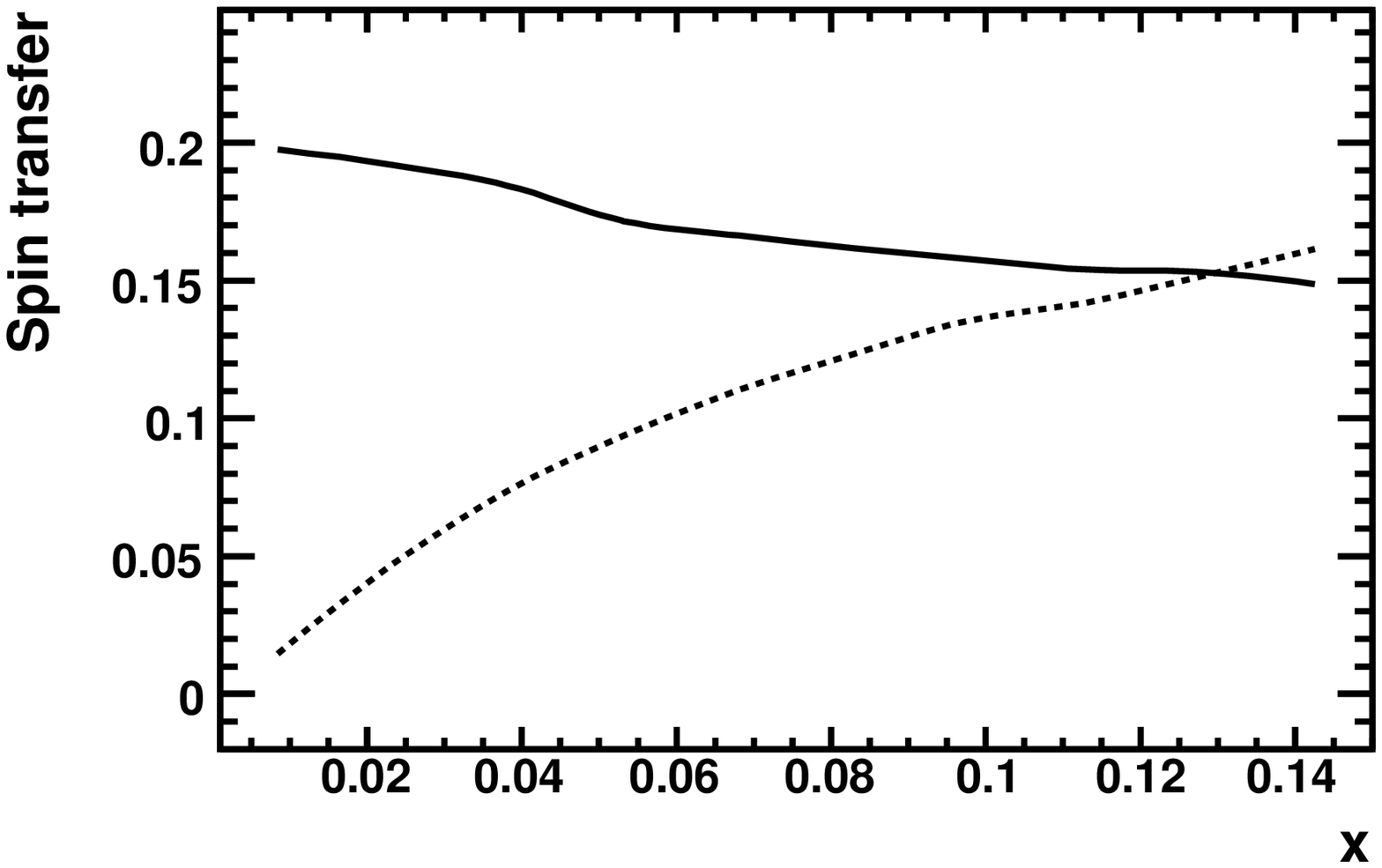,width=0.45\linewidth} &
\epsfig{file=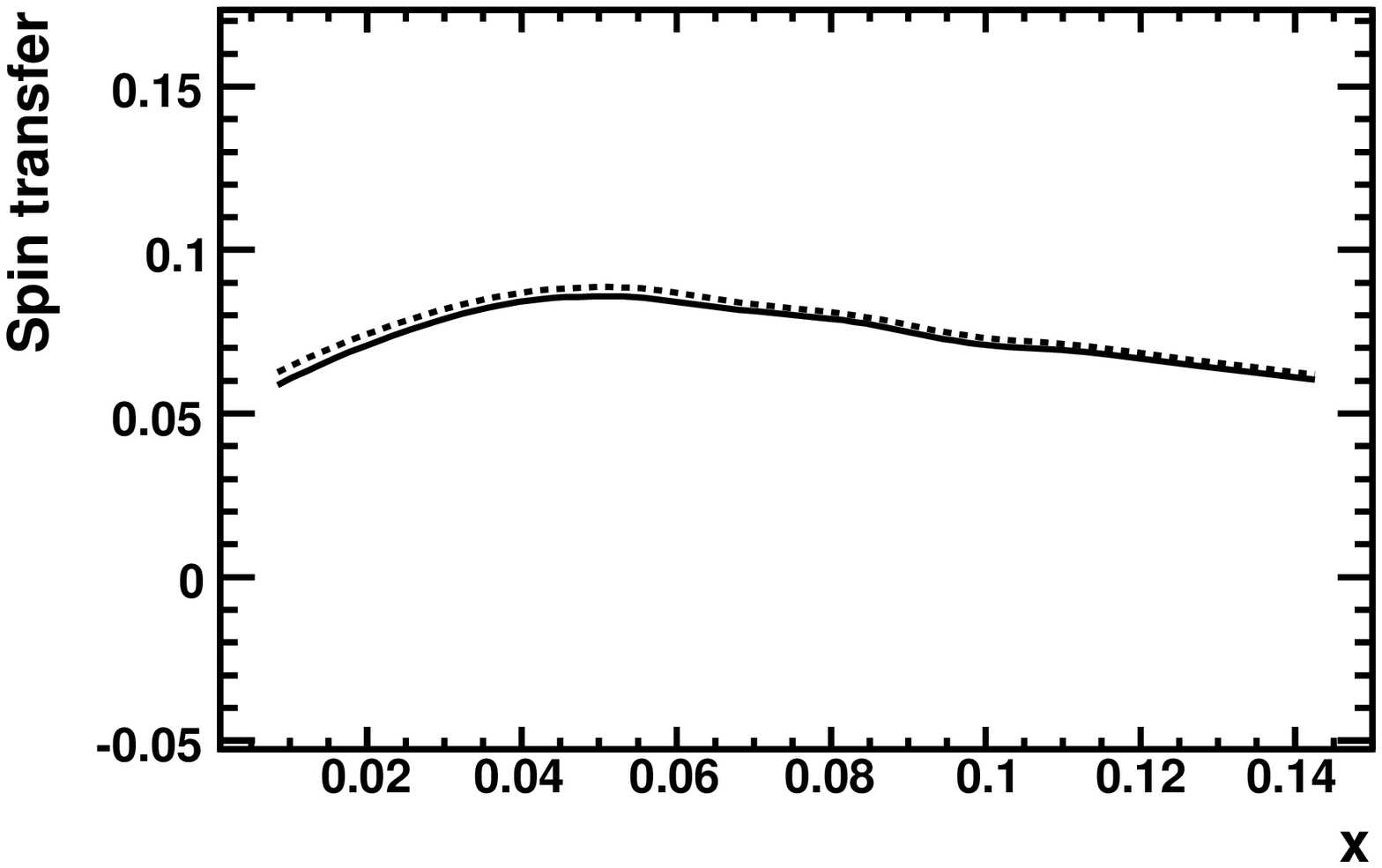,width=0.45\linewidth} \\
\epsfig{file=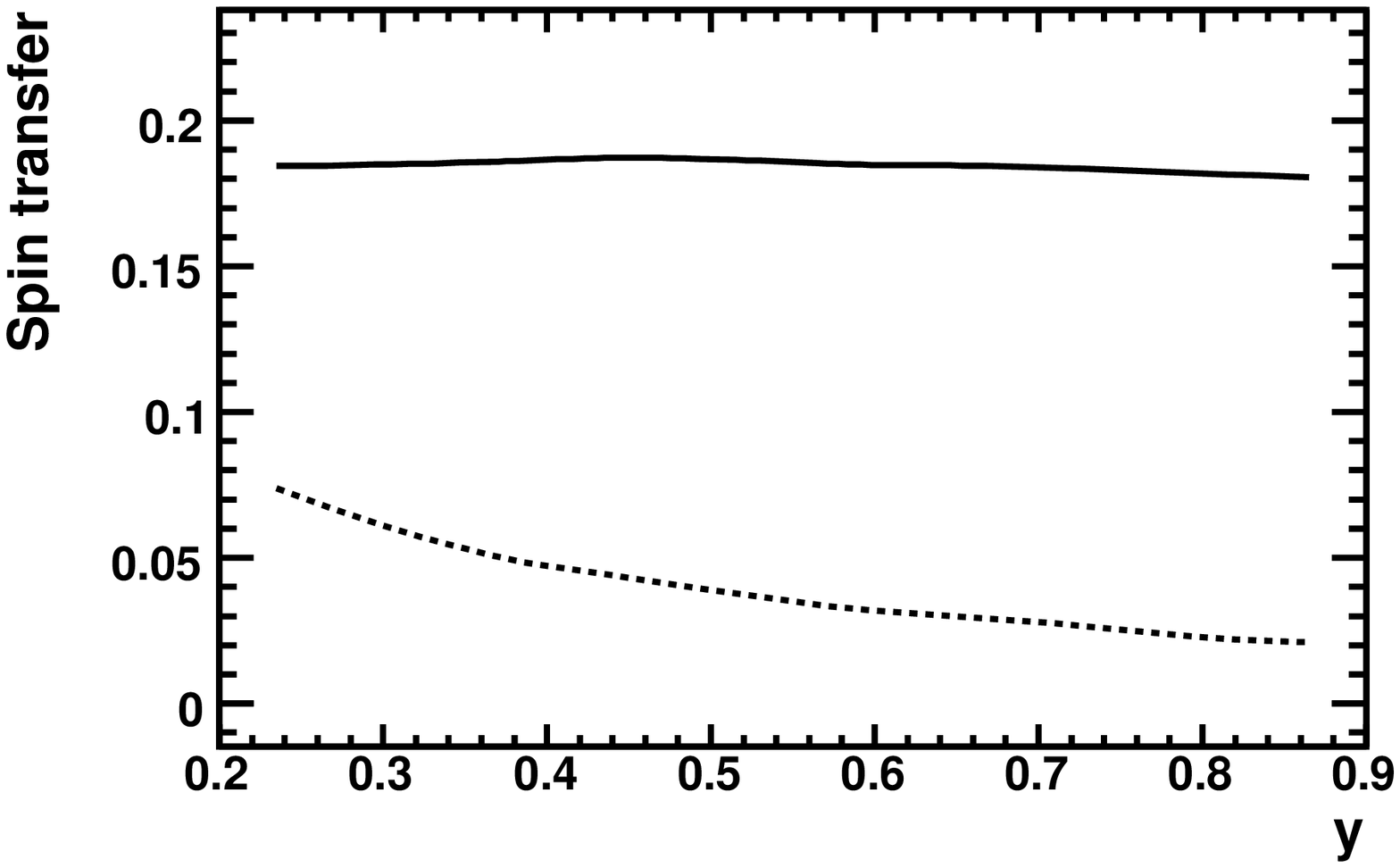,width=0.45\linewidth} &
\epsfig{file=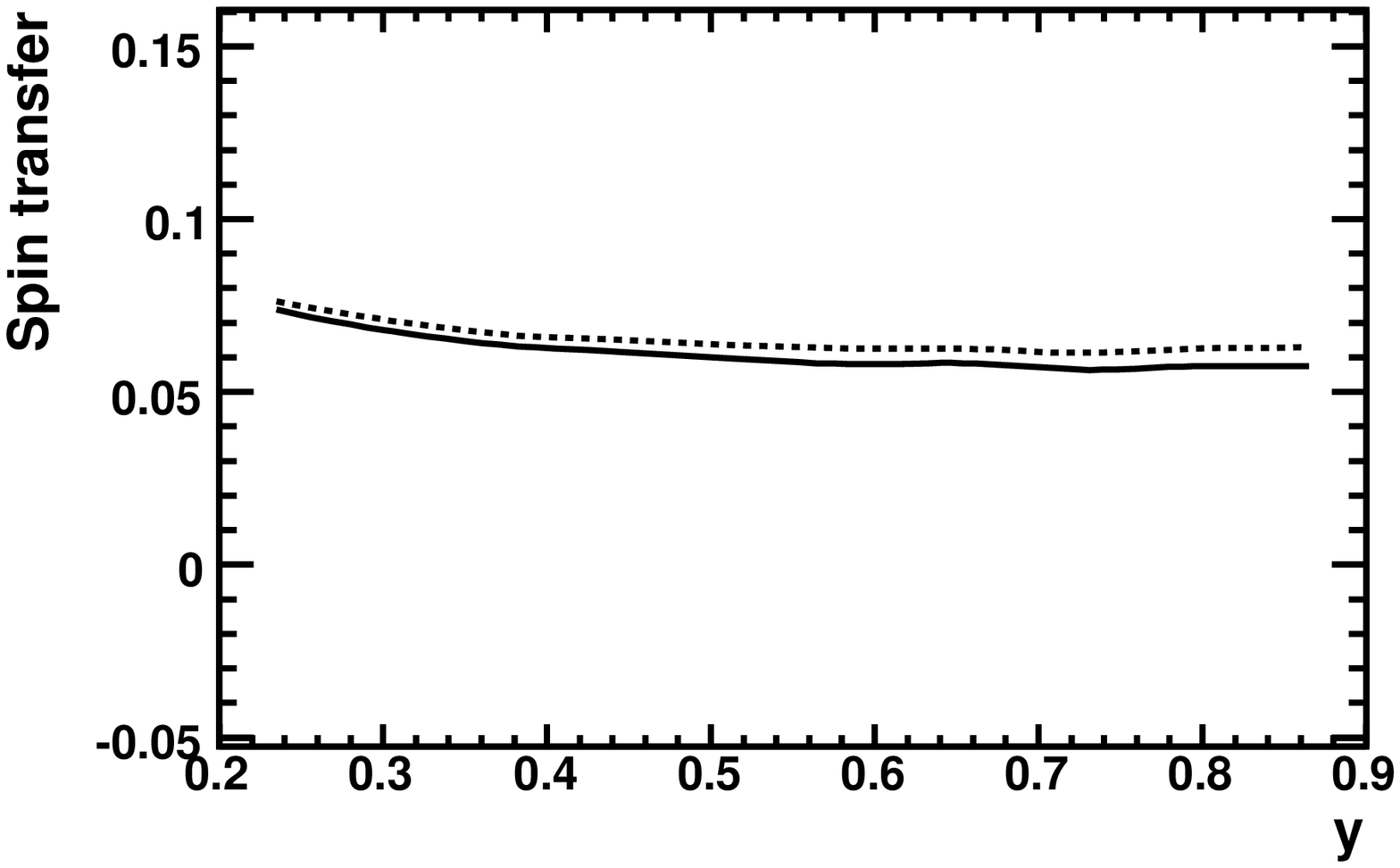,width=0.45\linewidth} \\
\end{tabular}
\caption{\label{fig:SpinTansferModelsAB} Comparison of the
spin-transfer predictions for Model A (solid line) and Model B
(dashed line) for $\lam$ (left) and $\alam$ (right) hyperons at
COMPASS energy
as functions of $x$, $x_F$ and $y$. We use the GRV98 parton
distribution functions and the SU(6) model of the baryon spin
structure.}
\end{figure*}
\begin{figure*}[htb]
\begin{tabular}{cc}
\epsfig{file=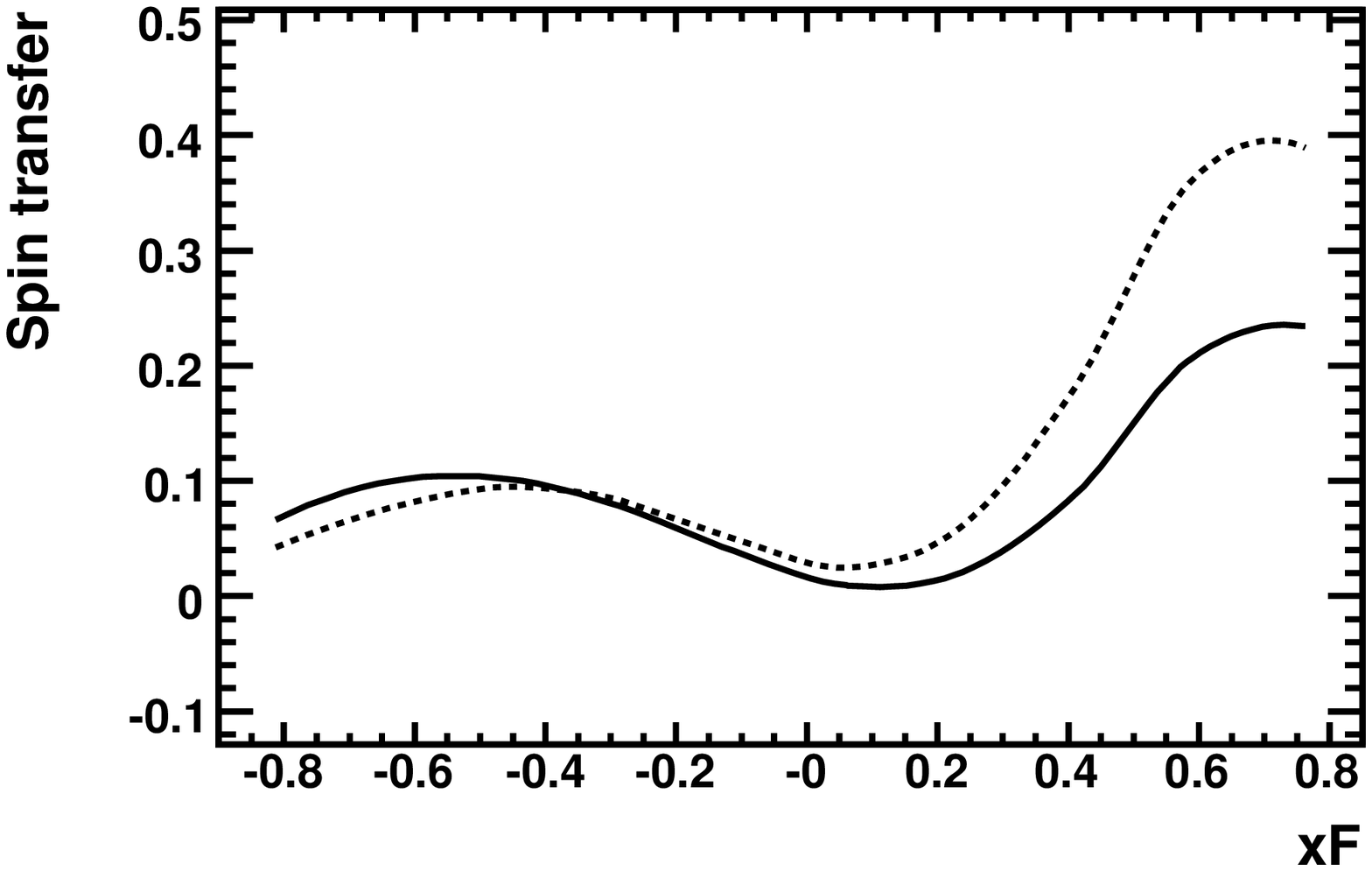,width=0.45\linewidth}&
\epsfig{file=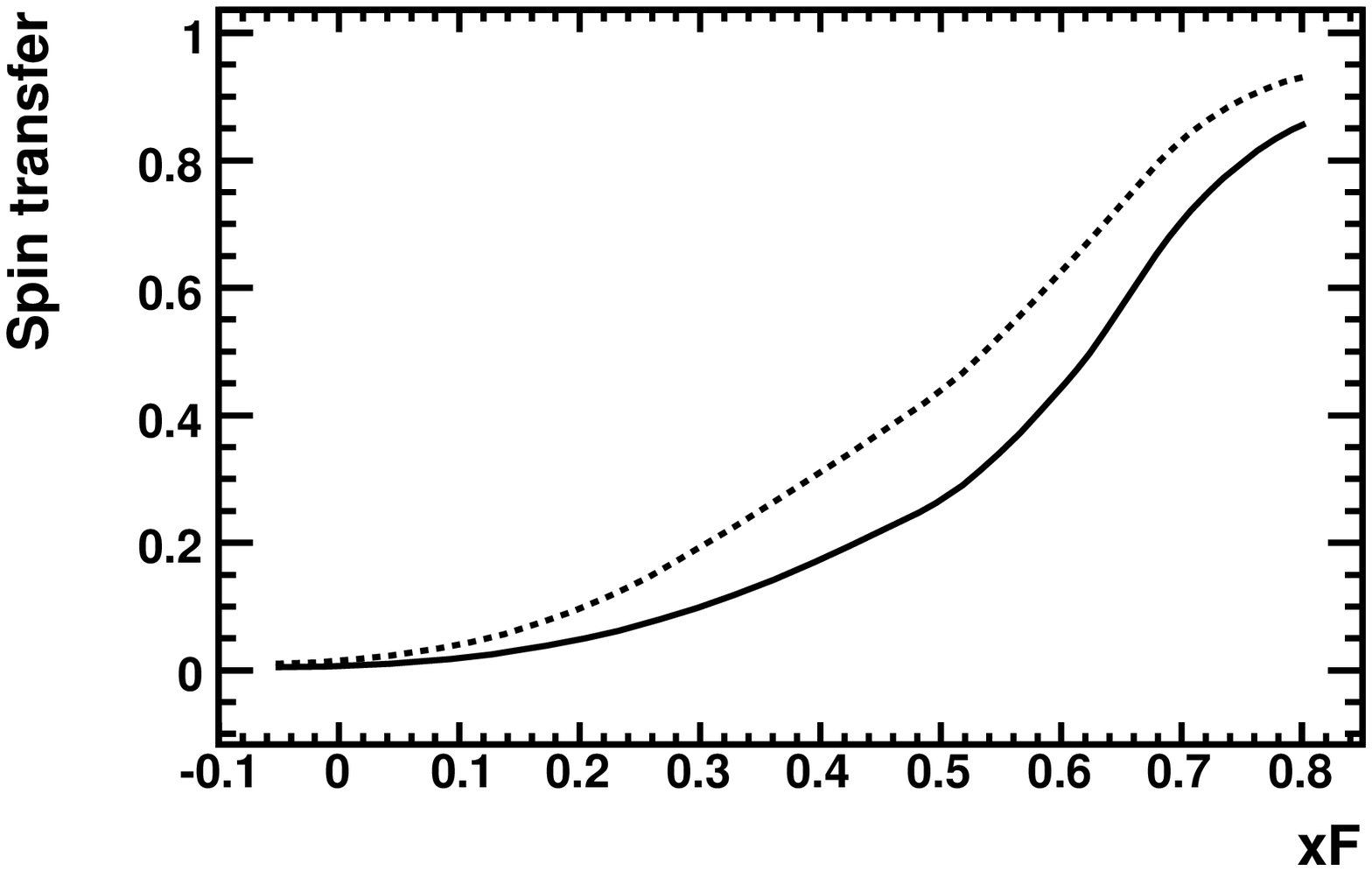,width=0.45\linewidth}\\
\end{tabular}
\caption{\label{fig:cteq} The spin transfers to the $\lam$ (left)
and $\alam$ (right) hyperons in the SU(6) model for the GRV98
(solid line) and CTEQ5L (dashed line) sets  of parton
distributions, as functions of $x_F$, at COMPASS energy.}
\end{figure*}
%

%

%
%
\begin{figure*}[htb]
\begin{tabular}{cc}
\epsfig{file=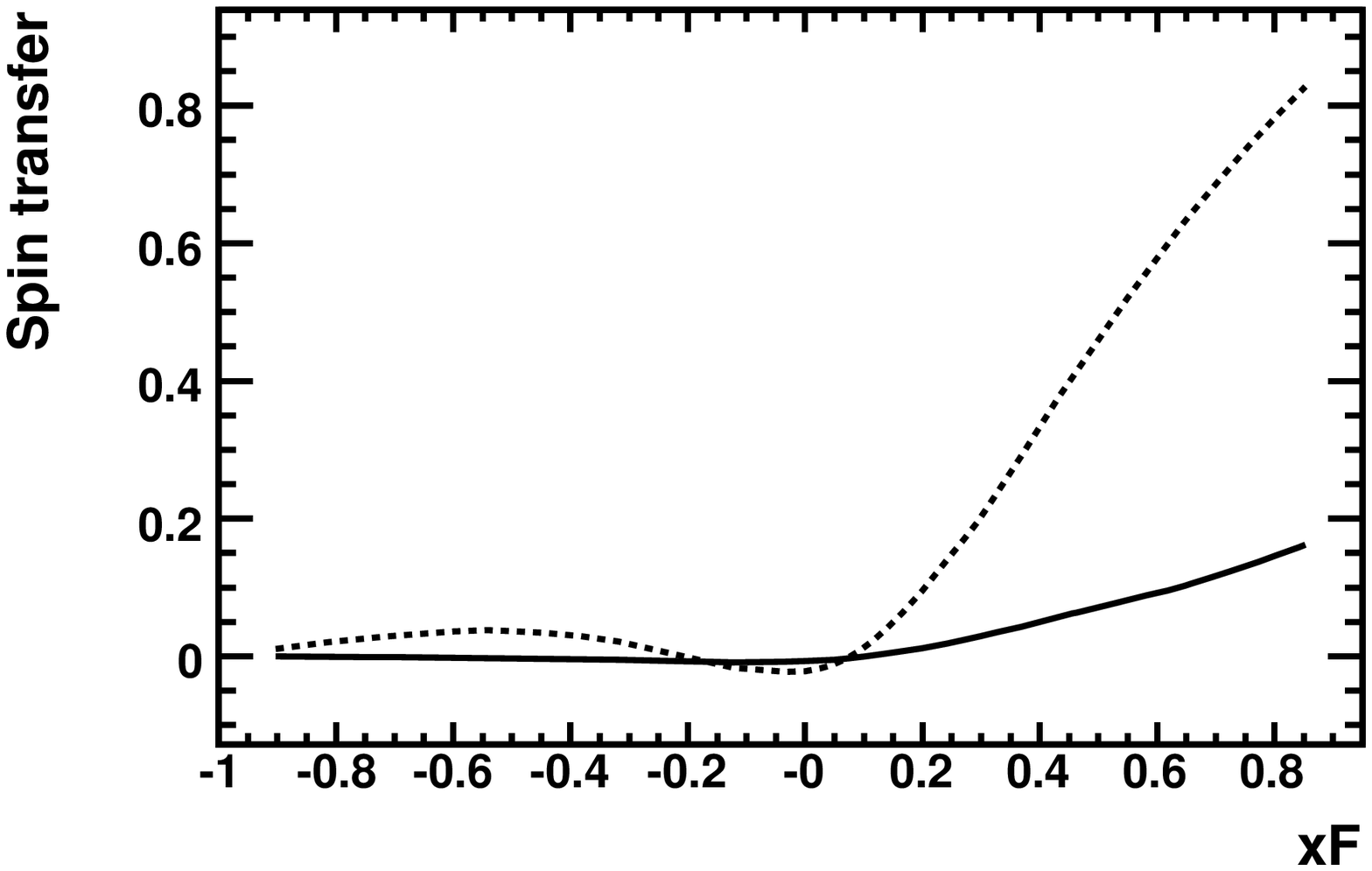,width=0.45\linewidth}&
\epsfig{file=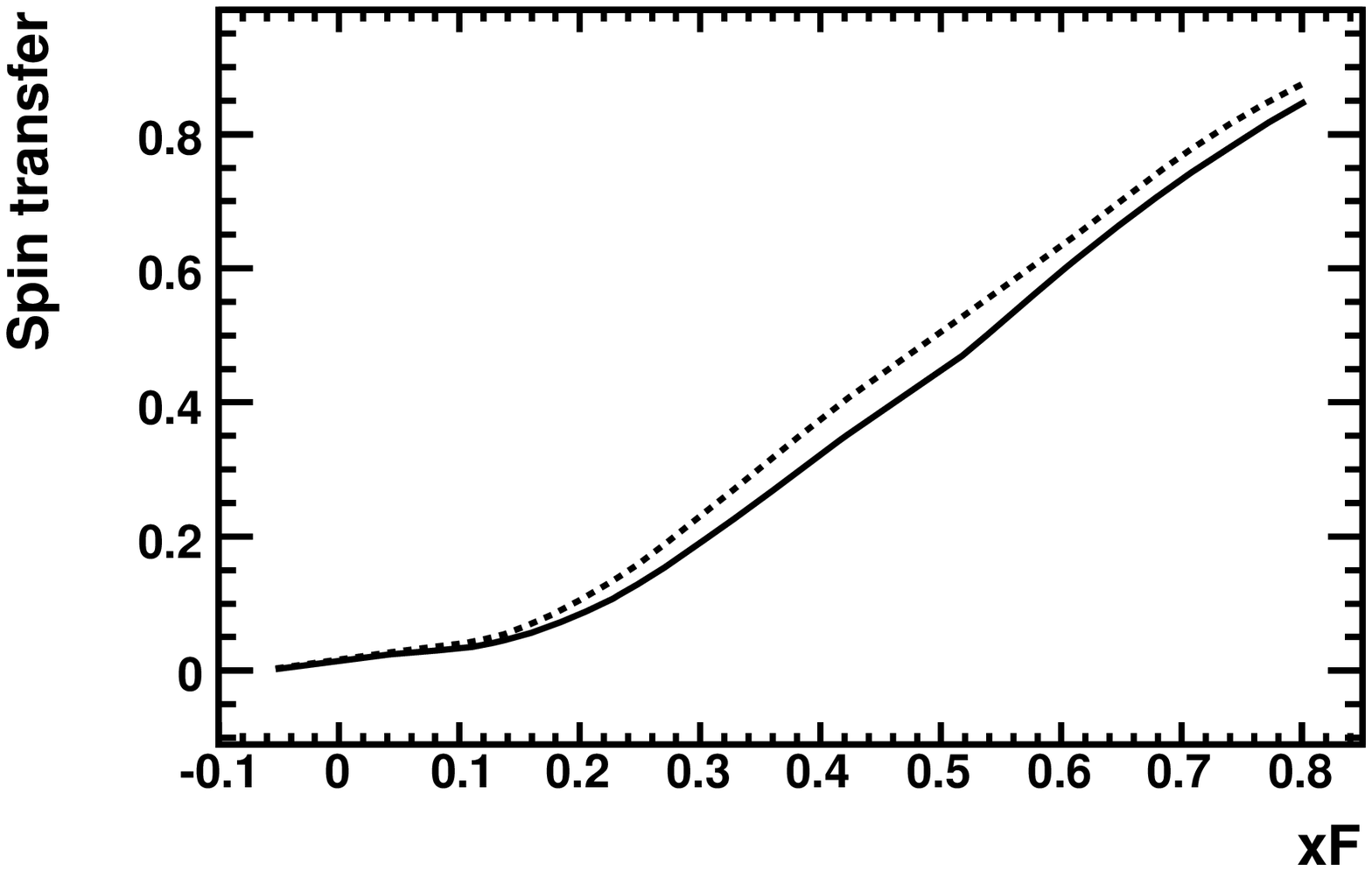,width=0.45\linewidth}\\
\end{tabular}
\caption{\label{fig:compare_grv98_cteq5l_su6_hera} The spin
transfer to the $\lam$ (left) and $\alam$ (right)  hyperons in the
SU(6) model for the GRV98 (solid line) and CTEQ5L (dashed line)
sets of parton distributions, as functions $x_F$, at HERA energy.}
\end{figure*}
%
%
\begin{figure*}[htb]
\begin{tabular}{cc}
\epsfig{file=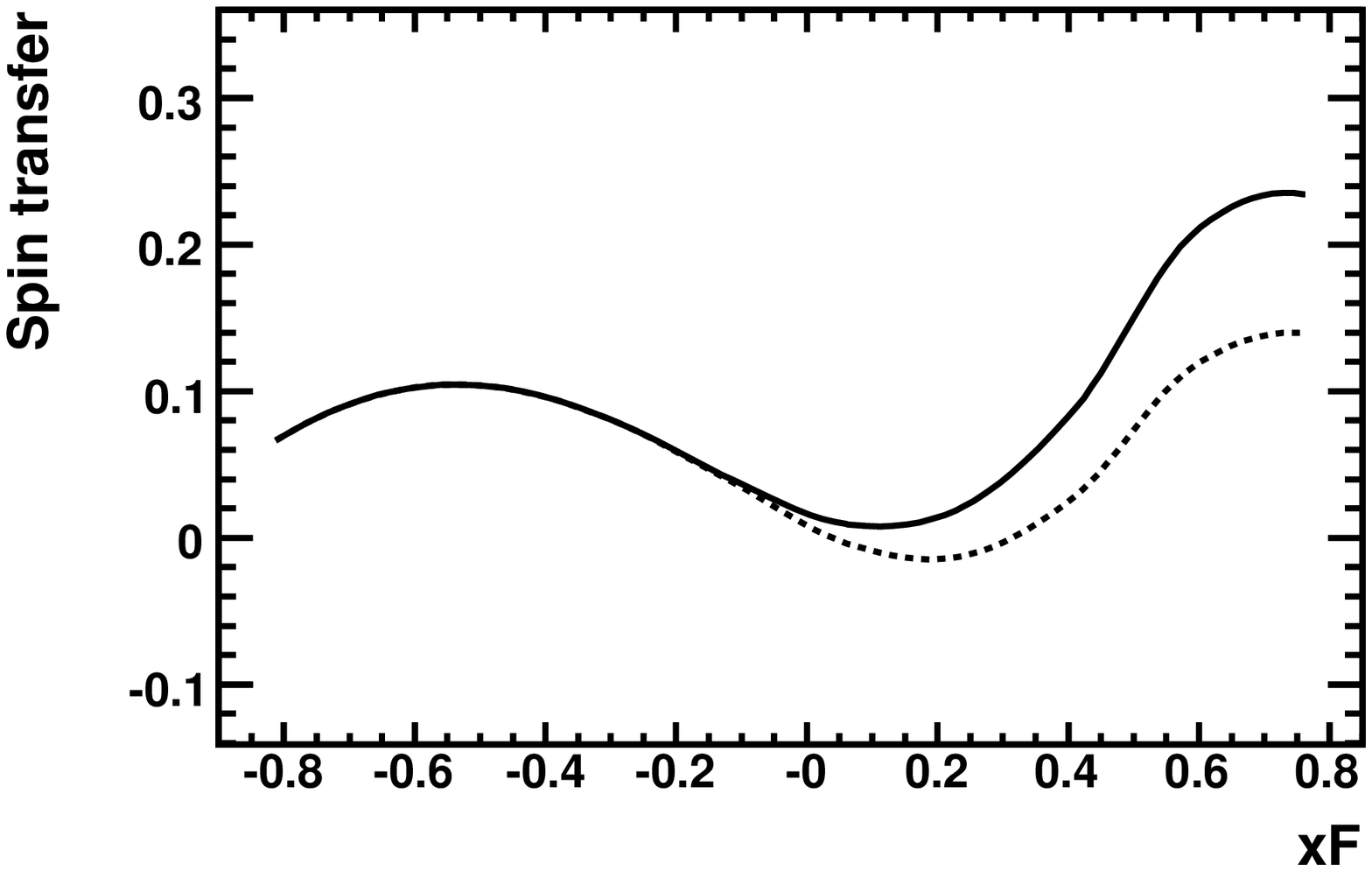,width=0.45\linewidth}&
\epsfig{file=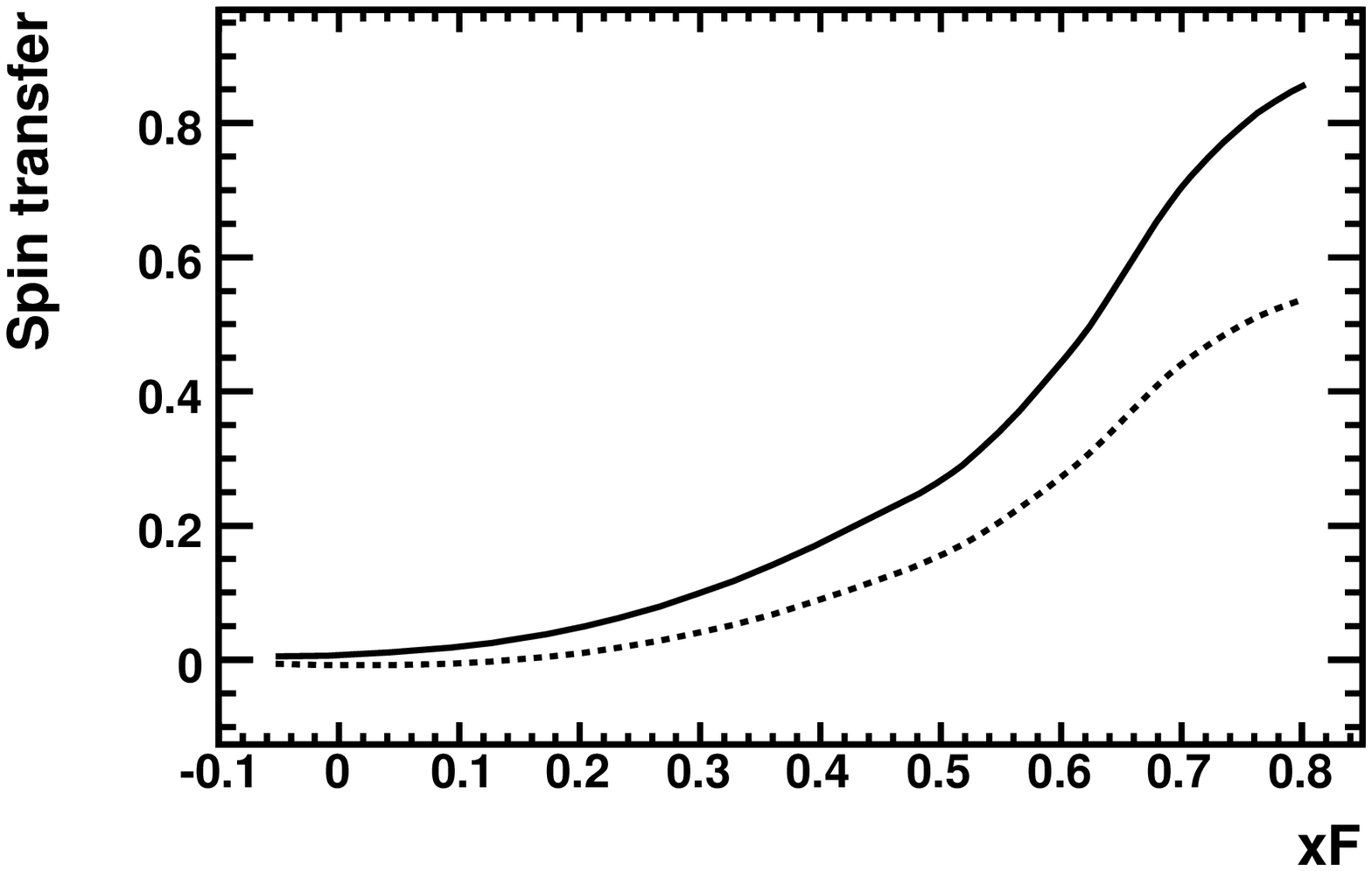,width=0.45\linewidth} \\
\end{tabular}
\caption{\label{fig:su6-bj} The spin transfers to the $\lam$
(left) and $\alam$ (right) hyperons in the SU(6) (solid line) and
BJ (dashed line) models, assuming GRV98 parton distribution
functions, as functions of $x_F$, at the COMPASS energy.}
\end{figure*}
\begin{figure*}[htb]
\begin{tabular}{cc}
\epsfig{file=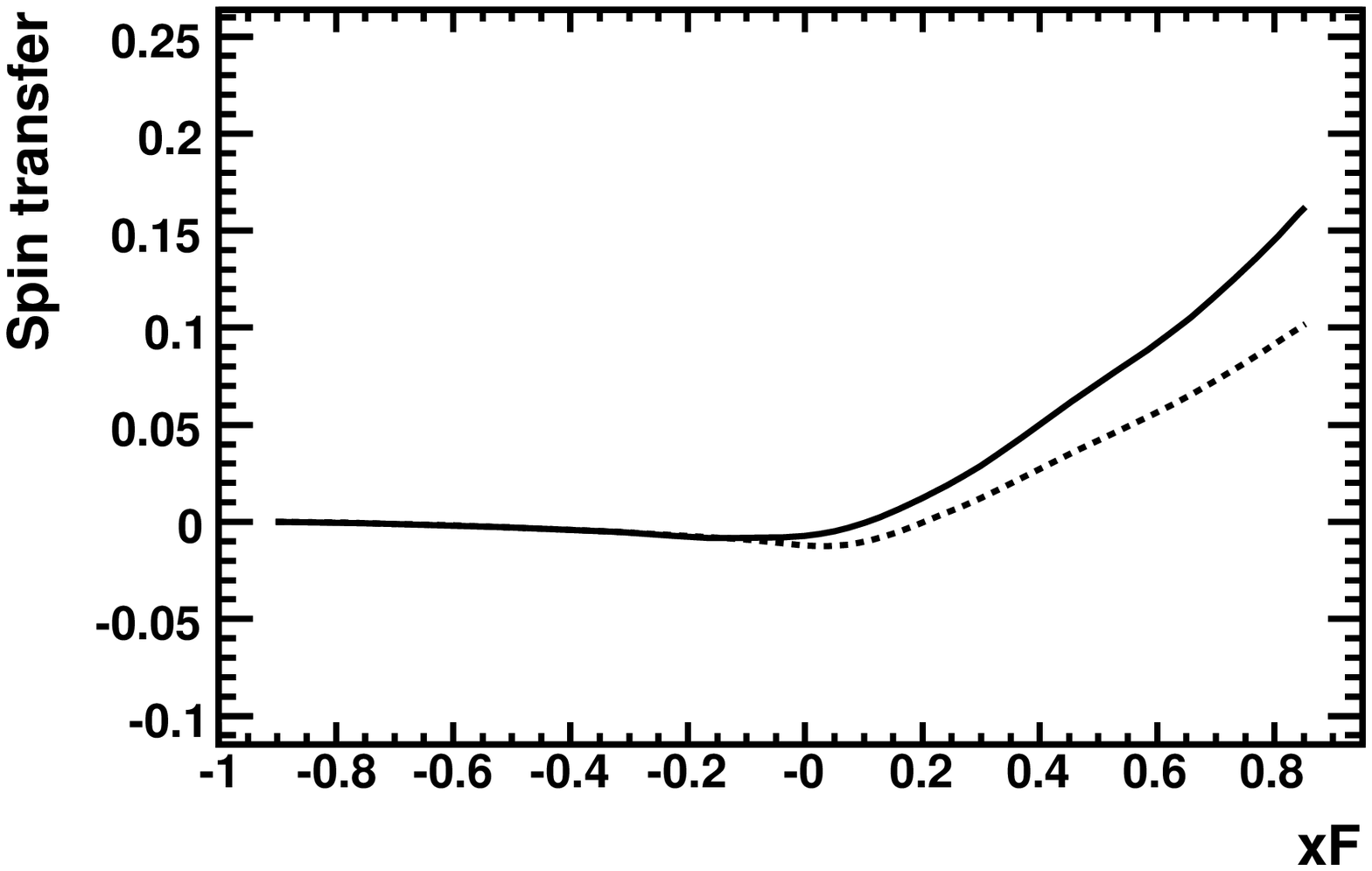,width=0.45\linewidth}&
\epsfig{file=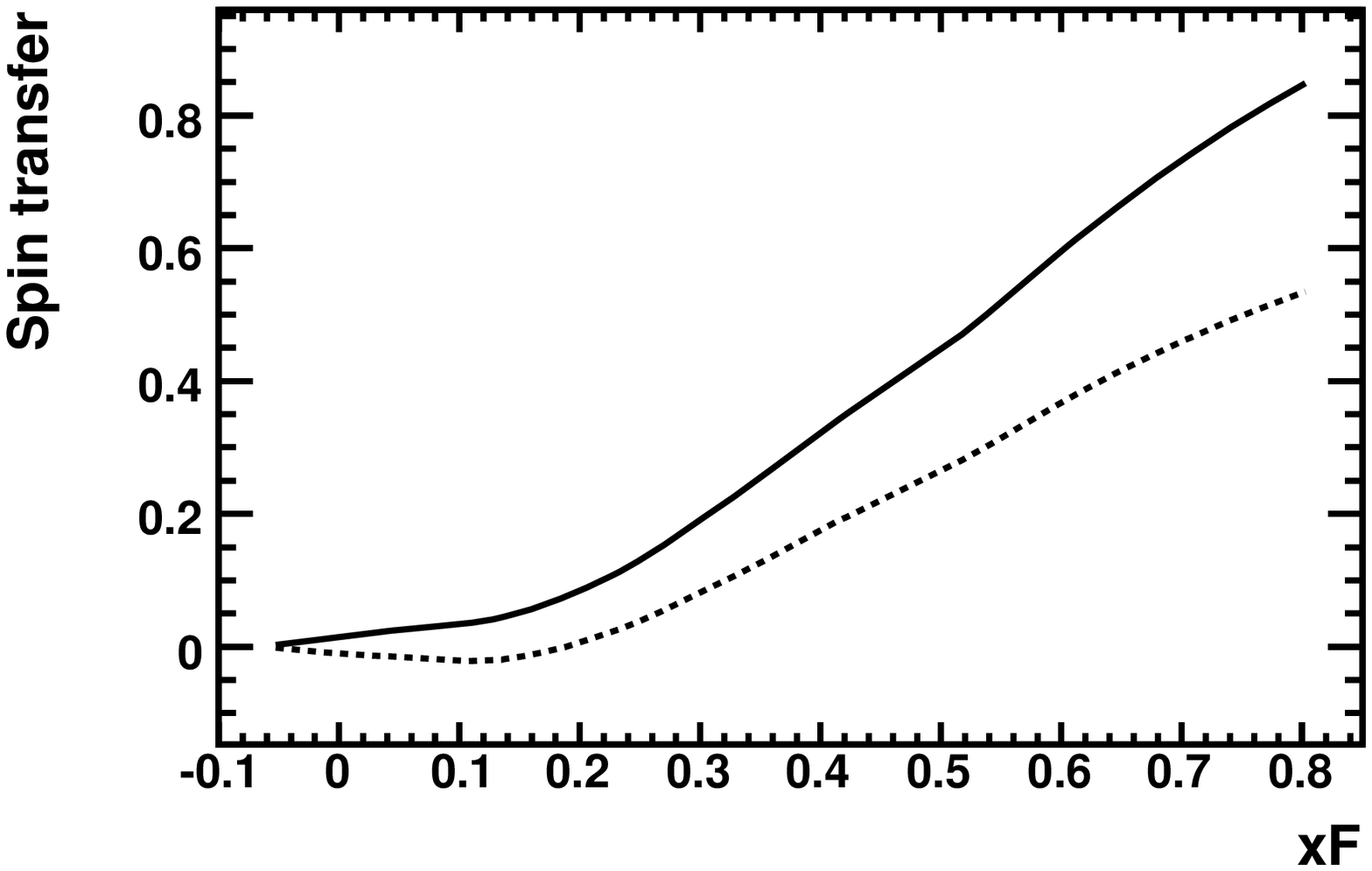,width=0.45\linewidth} \\
\end{tabular}
\caption{\label{fig:su6-bj-hera} The spin transfers to the $\lam$
(left) and $\alam$ (right) hyperons in the SU(6) (solid line) and
BJ (dashed line) models, assuming GRV98 parton distribution
functions, as functions of $x_F$, at the HERA energy.}
\end{figure*}
%
%
\begin{figure*}[htb]
\begin{tabular}{cc}
\epsfig{file=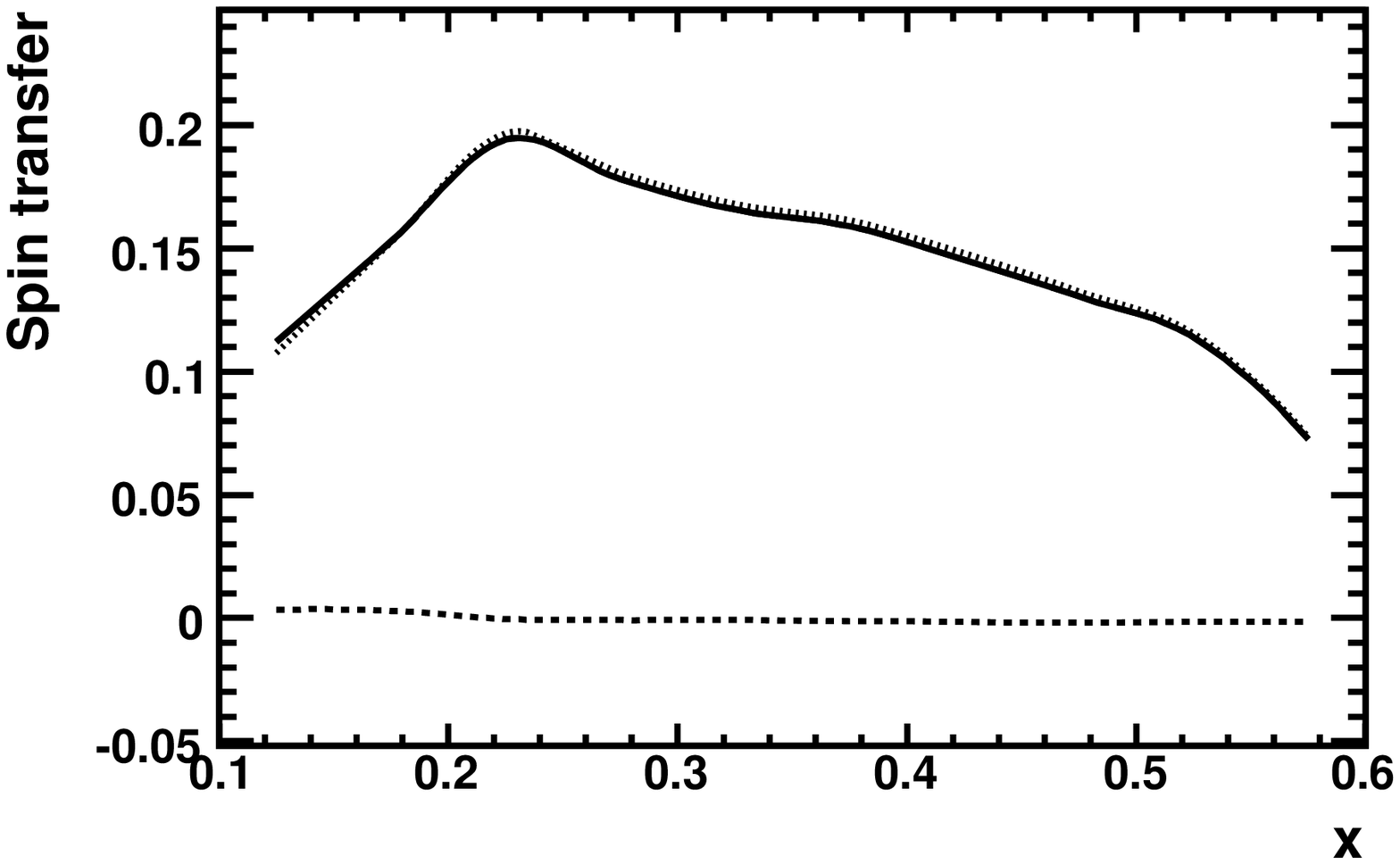,width=0.45\linewidth}&
\epsfig{file=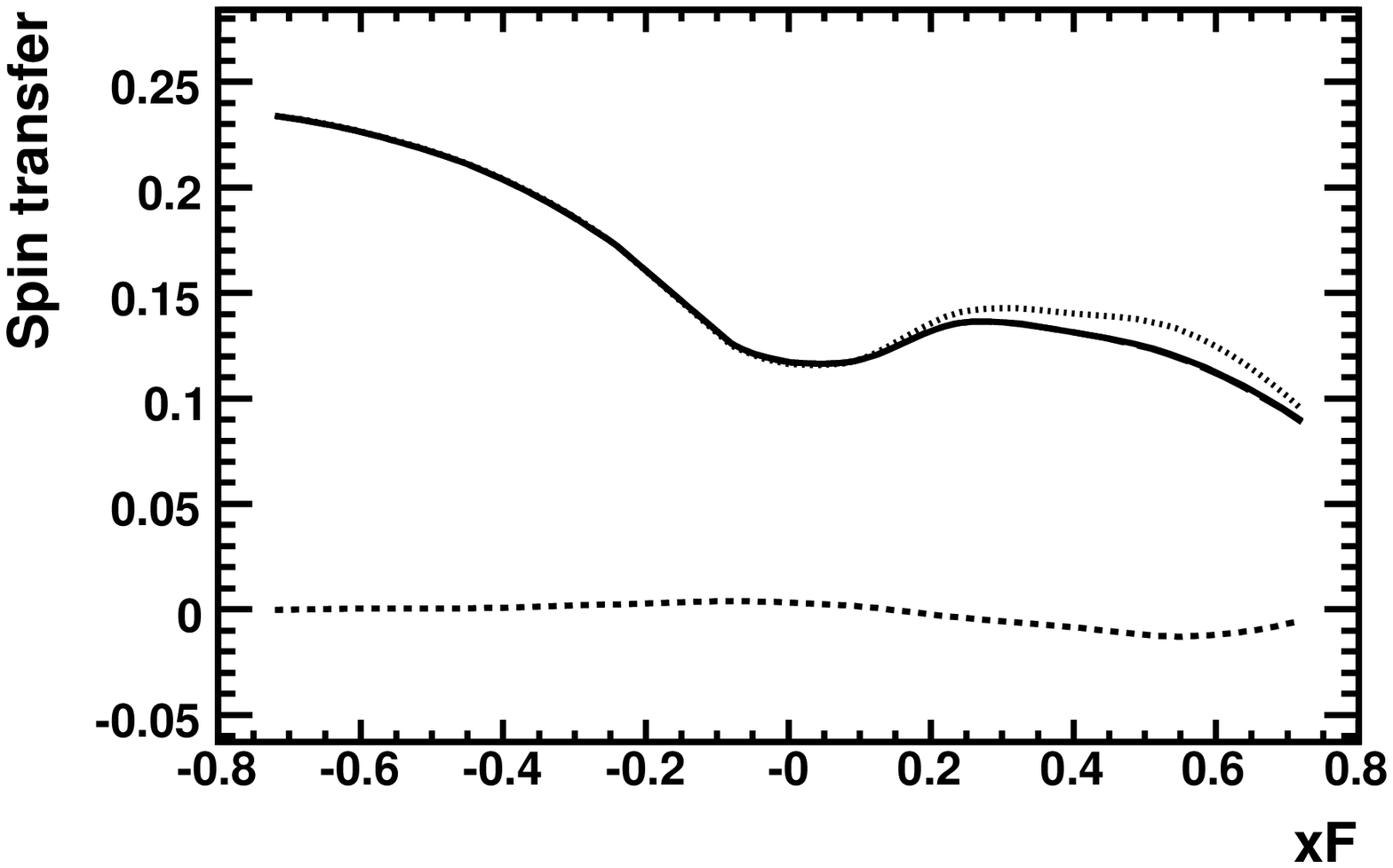,width=0.45\linewidth}\\
\end{tabular}
\caption{\label{fig:su6_grv98_jlab} The spin transfer to the
$\lam$  hyperons produced under JLAB experimental conditions in
the SU(6) model for the GRV98 set  of parton distributions, as
functions of $x$ (left) and $x_F$ (right). The solid lines
correspond to the full calculations, the spin transfers without
the contributions from $u$ and $d$ quarks are shown by thin dashed
lines, the spin transfers without the contributions from the $s$
struck quarks are shown by the dash-dotted lines, and the bold
dashed lines correspond to calculations with $C_{sq}=0$. }
\end{figure*}